\documentclass[journal=jctcce,manuscript=article]{achemso}
\usepackage{caption,float,geometry,natbib,setspace,xkeyval}
\usepackage{graphicx,subcaption,multirow,amsmath}
\title{A model Exact Study of the properties of low-lying electronic states of Perylene and Substituted Perylenes }
\author{Geetanjali Giri} 
\email{geetgiri1@gmail.com}
\affiliation[]{Solid State and Structural Chemistry Unit, Indian Institute of Science, Bangalore 560012, India}
\author{Suryoday Prodhan} 
\email{suryodayp@gmail.com}
\affiliation[]{Universit\'e de Mons-Hainaut, Place du Parc 20, 7000 Mons, Belgium}
\author{Y. Anusooya Pati}
\email{anuspati@gmail.com}
\affiliation[]{Solid State and Structural Chemistry Unit, Indian Institute of Science, Bangalore 560012, India}
\author{S. Ramasesha}
\email{ramasesh@iisc.ac.in}
\affiliation[]{Solid State and Structural Chemistry Unit, Indian Institute of Science, Bangalore 560012, India}
\date{\today}

\begin{document}
\begin{abstract}
 {
There is a resurgence of interest in the electronic structure of perylene for its applications in molecular devices such as organic photovoltaics and organic light emitting diodes. In this study, we have obtained the low-lying singlet states of perylene by exactly solving the Parisar-Parr-Pople model Hamiltonian of this system with 20 sites and 20 electrons, in the VB basis where dimensionality is $\sim$ 5.92 billion. The triplet states of perylene are obtained using a DMRG scheme with symmetry adaptation. The one and two photon states are very close in energy $\sim$ 3.2 eV while the lowest triplet state is slightly below 1.6 eV indicating that perylene is a good candidate for singlet fission. To explore the tunability of the electronic states, we have studied donor-acceptor substituted perylenes. The two donors and two acceptors are substituted symmetrically either at the four bay sites or four peri sites. In all the bay substitution and one peri substitution at moderate D/A strength, the optical gap is lowered to about 2.8 eV. These molecules can be used as blue emitters. We have also reported bond orders in all the cases and perylene as well as substituted perylenes can be viewed as two weakly coupled naphthalene in the singlet states but in triplets these bonds tend to be comparable to other bonds in strength. The charge densities in substituted perylenes are mostly localized around the substitution sites in the ground state. The positive spin densities in triplets are concentrated around the peri and bay sites with the remaining sites having small spin densities of either sign.}
\end{abstract}
\maketitle
\section{Introduction}
{\indent In recent years materials science has been focused on advancement of
nanotechnology, based majorly on graphene sheets and similar 2D and quasi
one dimensional carbon structures such as nanoflakes, nanotubes and
 graphene nanoribbons. In this context Polycyclic Aromatic Hydrocarbons
have attracted a great deal of attention because of their potential as 
active materials in organic optoelectronic devices \citep{guo17} as well as in biological
 imaging \citep{heek13}. Planar PAHs like perylene prove to be a model structure for these
 low dimensional materials.
Perylene $C_{20}H_{12}$, a conjugated organic molecule is an important member of PAHs and is also parent material of rylene dyes. Graphene-like structure of perylene (and its imide derivatives) provides a platform to make 2D organic devices  with high carrier mobility and improved efficiency \cite{guo17, schill18}. Perylene diimide (PDI) derivatives functionalized at the ortho-position (αPPID, αPBDT) were synthesized and used as electron acceptors in non-fullerene organic photo voltaic cells by Zhao et al ~\citep{zhao08}.
Recently Lai et al ~\cite{lai15} have reported the successful growth of free-standing ultra thin nanosheets ($<$10 nm) of Perylene which provides impetus to the study and application of 2D nanostructures of organic semiconductors.}

{\indent Over the years, electronic structure of perylene has been extensively studied using many theoretical and computational methods. The crystal and molecular structure of perylene was first reported by Donaldson et al \citep{donaldson53} in 1953. The electronic structure of crystalline perylene including van der Walls interactions has been investigated within the framework of Density Functional Theory (DFT-D2 method) by Fedorov et al ~\cite{fedorov13}.
The description of potential energy surfaces for intermolecular motion in a dimer of perylene based dyes specially energetics of perylene bisimide (PBI) has been given by Walter ~\cite{walter17} using TD-DFT methods. 
These First principle methods use mean-field potentials, hence they lack the complete description of excited state properties, particularly the triplet state and the two-photon states.
Since electron correlation effects are too strong to predict the correct energy spectra without extensive configuration interaction, DFT based methods can not reliably describe the excited states. It has been shown that a complete CI calculation within the $\pi$-framework based on the Pariser-Parr-Pople (PPP) model Hamiltonian with transferable parameters can accurately reproduce many of the experimentally studied properties of low-lying states of organic molecules. The conjugated $\pi$-system of perylene consists of 20 electrons delocalized over 20 $2p_z$-orbitals as carbon sites. Its geometrical structure can be viewed as two naphthalene molecules linked by two long bonds. In the present paper, we have studied model exact correlated electronic singlet states of unsubstituted perylene by exactly solving the PPP model using a diagrammatic valence bond (DVB) method.}\\


\begin{figure}[]
\begin{center}
\includegraphics[scale=0.5]{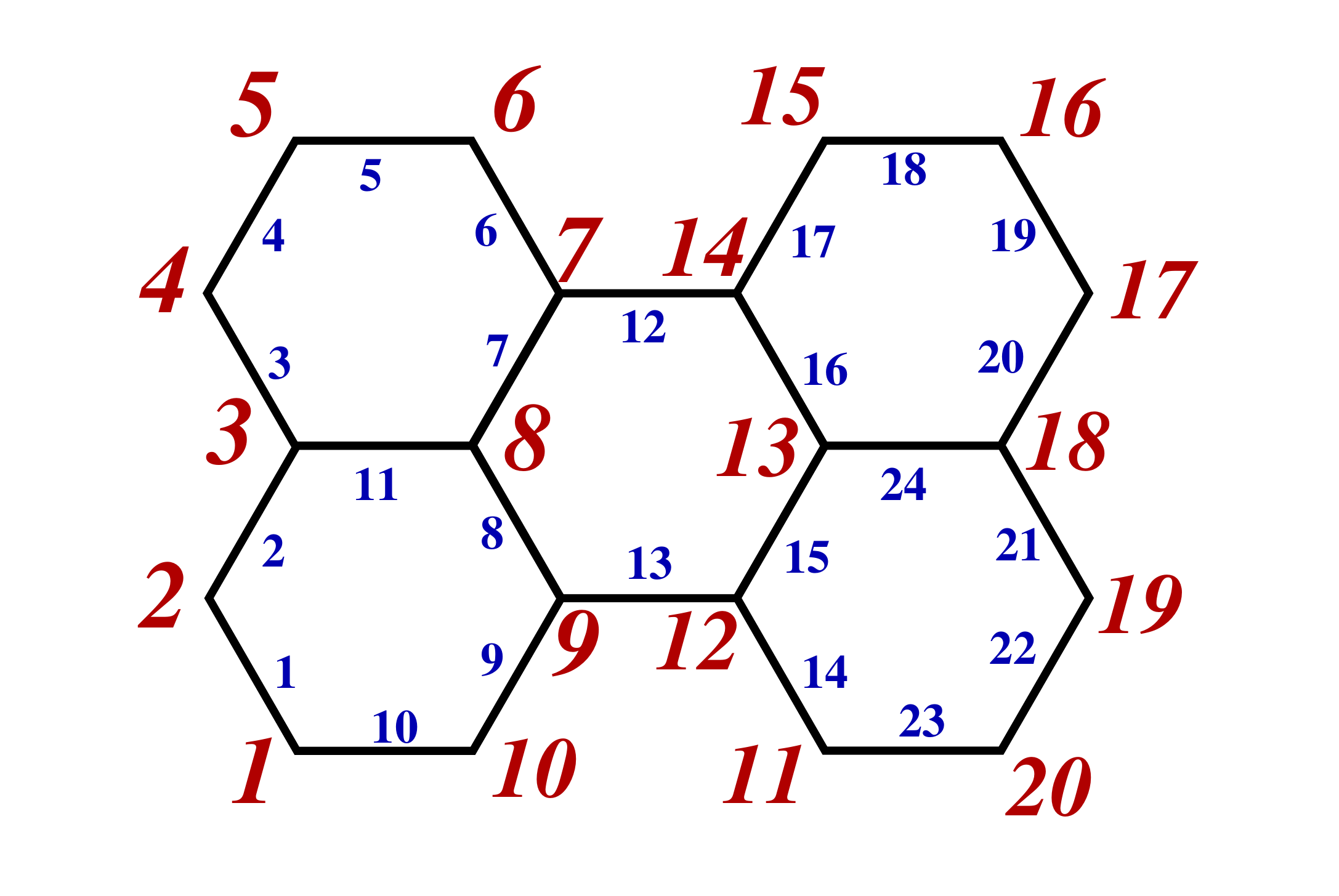}
\end{center}
\caption{Structure of Perylene molecule. Site numbers are shown in red outside the molecule and bond orders are shown in blue inside the molecule. Site numbers 2, 4, 17 and 19 are the peri sites and site nos. 6, 10, 11 and 15 are the bay sites.}
\label{fig: perylene}
\end{figure}

{\indent The full $\pi$-electron configuration space of perylene consists of about 5.9 billion singlets and over 12 billion triplets. Unsubstituted perylene belongs to the $D_2h$ point group and has approximate electron-hole symmetry. We can obtain exact energy levels of singlets by exploiting these symmetries. The triplet Hilbert space is too large for exact studies even after exploiting all the symmetries. Hence we employ the DMRG method for studying the triplet states. Substituted perylenes do not have electron-hole symmetry and may have spatial symmetry based on the nature of substitution. We have studied symmetrically substituted perylene at the peri or bay sites in Fig \ref{fig: perylene} using the DMRG method. Besides excited state energies, we have also characterized the states by bond orders, spin densities in the triplet states, spin-spin and charge-charge correlations and transition dipoles  between ground and excited states. We have examined the suitability of these systems for singlet fission.} 
 
{In the next section we discuss the model and the techniques we have employed for solving the models. In the third section we present results on unsubstituted perylene. In section four we discuss results on substituted perylenes. We end with a section on summary and conclusions.}\\
\section{Model Hamiltonian and Methodology}

\subsection{Parisar-Parr-Pople Hamiltonian and parameters}
{The Parisar-Parr-Pople model \cite{pariser53,pople55} hamiltonian is defined over the carbon $2p_z$ orbitals involved in $\pi$-conjugation and includes explicitly the on-site as well as intersite electron-electron interactions. PPP model consists of two parts; one the non-interacting part which is just the H\"{u}ckel model with transfer integrals and site energies and the other that includes electron-electron interactions. The electron-electron repulsion integrals are treated within a zero differential overlap (ZDO) approach. The Hamiltonian can be written as}
\begin{align}
\hat{H}_{PPP}=\sum_{<ij>,\sigma} t_{ij} (\hat{c}^\dagger_{i,\sigma} \hat{c}_{j,\sigma} + H.C.) + \sum_{i} \epsilon_{i}\hat{n}_i
\mspace{48mu}
\notag\\
 + \sum_{i}\dfrac{U_i}{2}\hat{n}_i(\hat{n}_i-1) + \sum_{i>j} V_{ij}(\hat{n}_i - z_i)(\hat{n}_j - z_j)
\end{align}
{\noindent Here $t_{ij}$ is nearest-neighbour transfer term between two bonded sites `i' and `j'; $\hat{c}^\dagger_{i,\sigma}$ ($\hat{c}_{i,\sigma}$) creates (annihilates) an electron of spin $\sigma$ in $2p_z$ orbital of carbon atom at site i; $\epsilon_i$ is the site energy of `i'-th carbon atom. These constitute the non-interacting part of the PPP Hamiltonian. The interacting part of Hamiltonian consists of $U_i$, the on-site Coulomb repulsion term and $V_{ij}$s which are intersite electron repulsion integral between sites i and j. The $V_{ij}$s are obtained by interpolating between $U_{ij}$ and $\dfrac{e^2}{r_{ij}}$ as $r_{ij} \rightarrow \infty$ using the Ohno scheme. The operator $\hat{n_{i}}$ are the number operator, the C-C bond length of 1.4 $\AA$ is assumed for nearest neighbour bonds. For carbon atoms in conjugation, the PPP model parameters are obtained  by semi-empirical methods, and the transferable parameters widely used are  $t_0= -2.4 eV$, $U= 11.26$ eV and site enregies $\epsilon=0.0$ eV for unsubstituted molecule. To introduce the effect of donor-acceptor substitution, non-zero site energies $\epsilon_{i}$ are used at the donor/acceptor sites. A positive $\epsilon_{i}$ implies a donor and a negative $\epsilon_{i}$ is an acceptor and the strength of substitution is dictated by the magnitude of $\epsilon_{i}$. The chemical potential $z_i$ is the number of electrons at site `i' which leaves the site neutral. In our calculations we have taken $z_i$ value equal to 1 corresponding to one electron per carbon atom at a neutral site.}

{The PPP model is a non relativistic model and hence conserves total spin as well as z-component of the total spin. Using Slater determinants as basis allows exploiting only total $S_z$ conservation and hence the basis will be huge. We can further factorize the basis by choosing basis function which have a definite total spin. For this we employ a diagrammatic valence bond (DVB) basis which is a complete set of linearly independent but non-orthogonal basis and conserves the total spin as well as z-component of total spin. We have considered the perylene molecule to be planar. Consequently, $\sigma_{h}$ is an identity and the spatial symmetry of perylene reduces to $D_2$ point group. The diagrammatic valance bond basis for 20 electrons in 20 orbitals has 5,924,217,936 singlet basis functions; the number of triplets goes up to  12,342,120,700. Taking into account the $C_2$ and alternancy or electron-hole symmetry in perylene, combination of singlet configurations within $A_g^+$ space is 1,481,162,738 and in the $B_u^-$ space it is 1,481,122,588.}

{To obtain the singlet energies of perylene molecule we first obtained symmetry adapted basis functions and generated PPP hamiltonian matrix elements. The resulting matrix is non-symmetric and sparse in nature. We then used Rettrup's method to obtain low-lying eigenstates of the matrix. These eigen vectors were transformed into slater basis for property calculations to avoid dealing with the nonorthogonal VB basis.} 

{While full symmetry adaptation for the perylene molecule under $D_2$ point group is possible within the VB scheme, the computational resources required are huge and out of reach at present \cite{ramasesha84,soos84}. This is because in the dual basis approach we need to set-up matrices of Hamiltonian and relevant symmetry operators in the Slater determinantal basis which spans a Hilbert space of over 34 billion (34,134,779,536) basis states. To recover this bottleneck we employed a single spatial symmetry for each set of calculations. Thus, we carried out two sets of calculations one with only $C_2(y)$ spatial symmetry and the other with only $C_2(z)$ symmetry. We cannot exploit the $C_2(x)$ symmetry as it results in `illegal' VB diagrams which are difficult to break up into `legal' VB diagrams ~\cite{soos93}. Using these symmetries separately, we have obtained two states in the covalent VB space and two in the ionic VB space. We have then computed the transition dipoles from the ground state to the eigenstates in the ionic VB space and from the polarization direction of the transition dipole, we have obtained the symmetry label of the state in the $D_{2h}$ point group. We have followed the same approach for the triplet states within the DMRG scheme.}\\
\subsection{Density Matrix Renormalization Group Method for Perylene }
\begin{figure}[]
\begin{center}
\subcaptionbox[]{}{\includegraphics[width=1.0\textwidth]{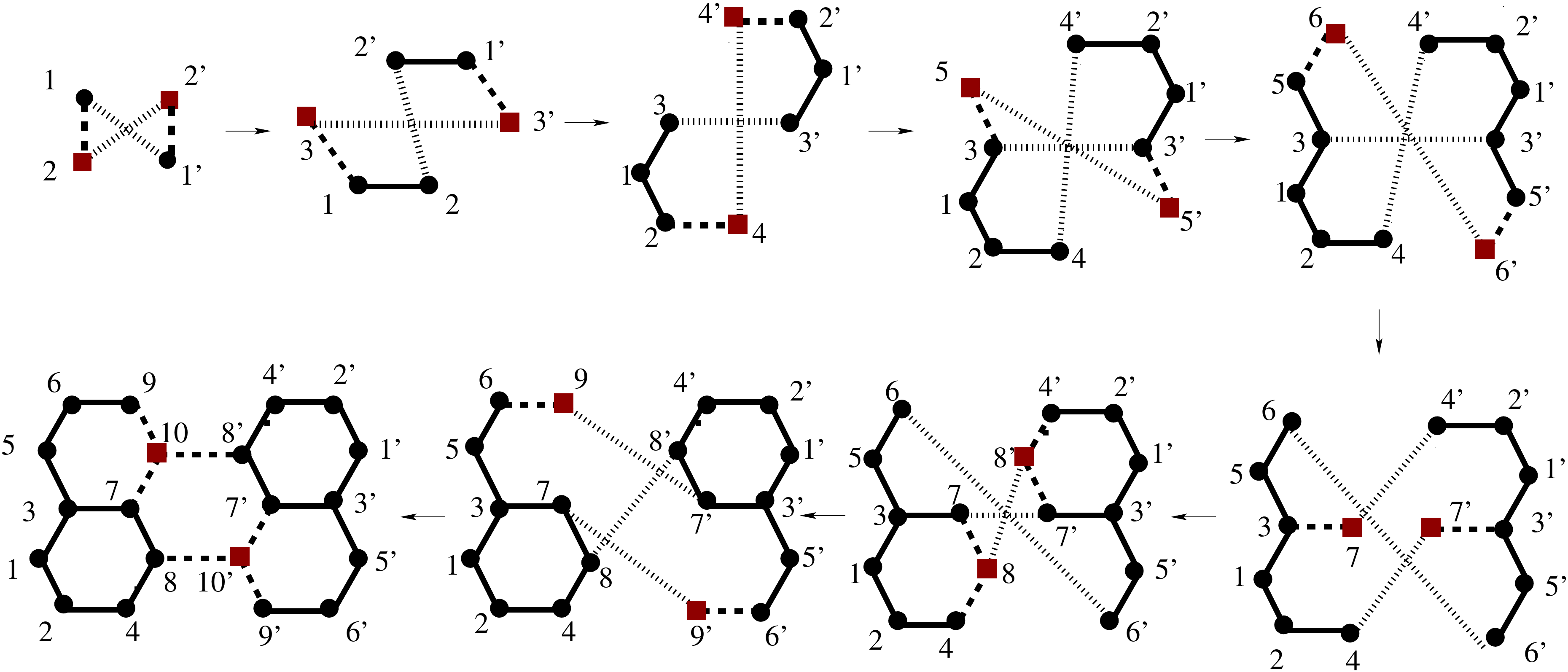}}%
\hfill \vspace{0.25cm}
\subcaptionbox[]{}{\includegraphics[width=1.0\textwidth]{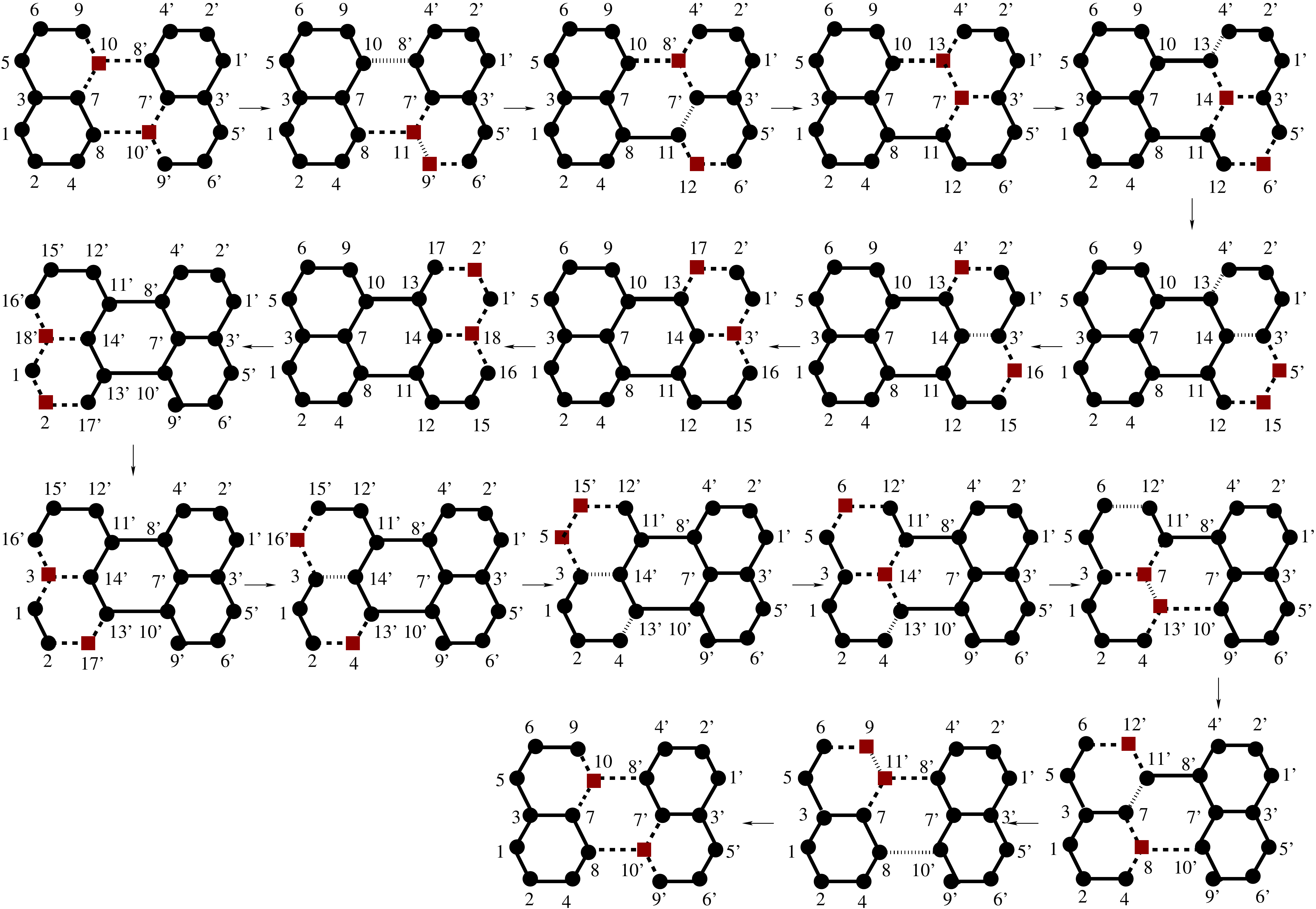}}%
\end{center}
\caption{Infinite (a) and Finite (b) DMRG Schemes for building the perylene molecule starting from 4 sites. Newly added sites are shown as red squares and old sites are shown as circles. Solid lines are bonds within the block, broken lines are bonds between old and new sites and dashed lines represent bonds between the blocks connecting old-old sites.}
\label{fig: pery-dmrg-scheme}
\end{figure}

\begin{figure}[]
\begin{center}
\includegraphics[scale=0.30]{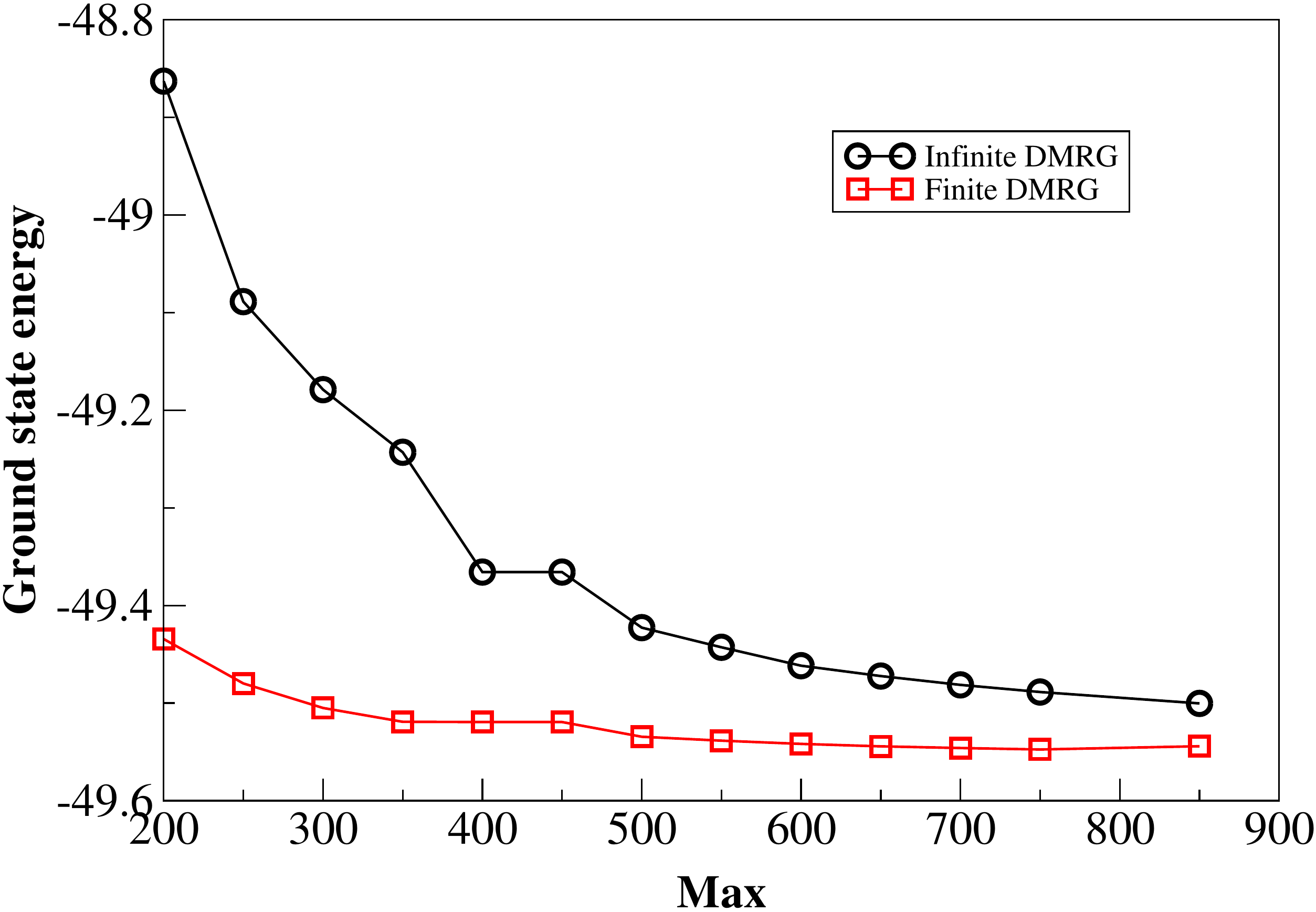}
\end{center}
\caption{Ground-state energy as a function of maximum no. of Density matrix eigen vectors kept at each iteration within PPP model.}
\label{fig: max-enr}
\end{figure}

\begin{figure}[]
\begin{center}
\subcaptionbox{Ground state}{\includegraphics[width=0.35\textwidth]{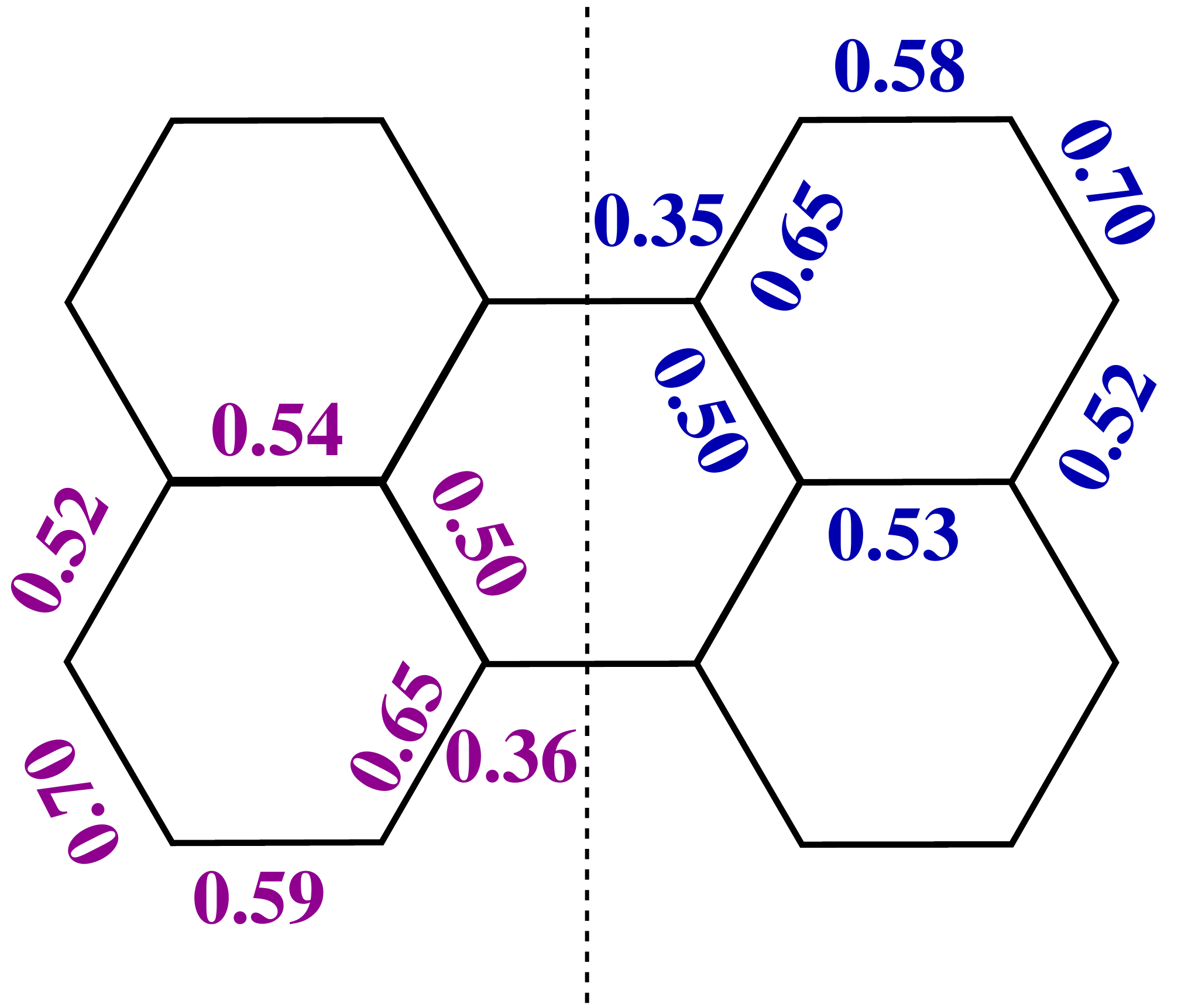}}%
\hspace{0.1cm} \vspace{0.2cm}
\subcaptionbox{Lowest dipole allowed state}{\includegraphics[width=0.35\textwidth]{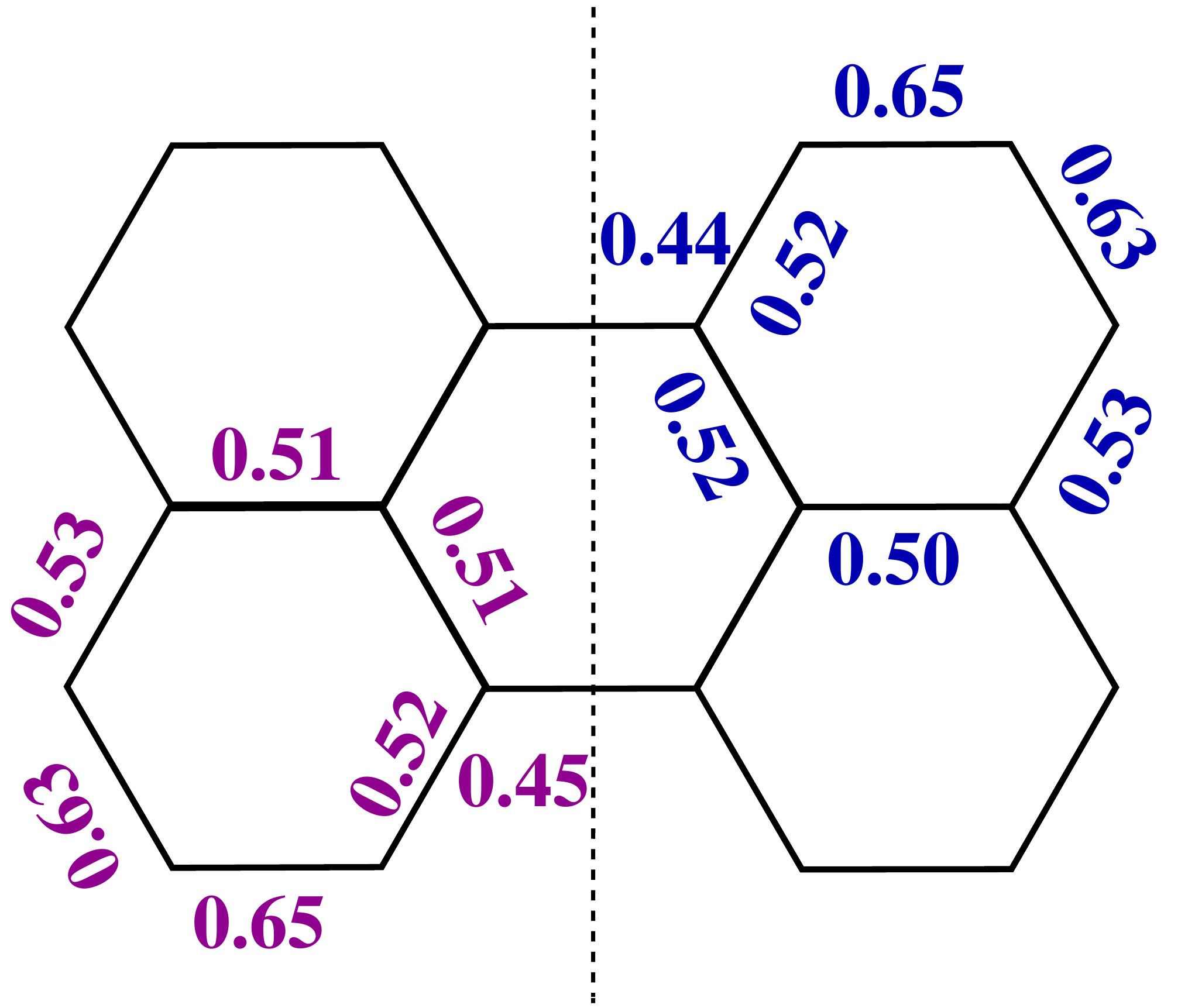}}%
\end{center}
\caption{Comparison of unique Bond orders using exact numerical method (left half, in magenta) and using DMRG method (right half, in blue).}
\label{fig: BO-unsub-s0}
\end{figure}

{ The DMRG algorithm, an iterative block-building scheme for calculation of low-lying eigen states of an interacting hamiltonian, truncates Hilbert space at every step keeping the number of block states which are density matrix eigenvectors, fixed \cite{PhysRevLett.69.2863,RevModPhys.77.259}. Thus the dimensionality of the truncated Hilbert space is kept almost a constant. The DMRG method is known to be very accurate for gapped one-dimensional systems as the entanglement entropy between two parts of the system remains independent of system size. Thus DMRG accuracy will be the same for all system sizes for the same cut-off in the number of block states. In our case, we have two factors which take the system away from one-dimensionality. The first is the model Hamiltonian. The intersite interactions in the PPP model renders the system dimension to be the order of the number of sites as each site interacts with every other site. However, in an earlier paper \cite{shaon12} we have shown that the entanglement entropy in the PPP model is not different from that in a Hubbard model and the reason being the intersite interactions are site diagonal and do not contribute significantly to entanglement. The second reason is perylene is not a one-dimensional molecule. However, the extension of the molecule in the two-dimensions in the planar molecule is small. Besides, the topology of the molecule is such that the largest transfer operator is only at most three neighbours away (see Fig~\ref{fig: pery-dmrg-scheme}). Therefore, it is possible to build the perylene molecule, such that the entanglement between the two parts of the molecule remains approximately the same and also reasonably small. This is achieved by suitably choosing the sites that are added at each step. Since we are also symmetrizing the scheme, the new sites are also added such that the electron-hole and the $C_2$ symmetry are retained at every stage of the infinite DMRG scheme \cite{suryo18}.}
\begin{table}[]
\caption{Comparison of energies from exact and DMRG calculations of different states of Perylene in the H\"{u}ckel and PPP models. The energies in eV are computed for t = -2.4 eV and U = 11.26 eV (in PPP model). The DMRG cut-off for number of block states kept is set to 800.}
\begin{center}
\setlength{\tabcolsep}{5pt}
\begin{tabular}{c c c c c}
\hline\hline 
Model & State/gap    &   Exact  &   DMRG  & Error   \\  \hline \vspace{0.2cm} 
H\"{u}ckel & $1 ^1A_g^+$  & -67.776 & -67.763 & 0.02 \%  \\ \vspace{0.2cm}
PPP    &   $1 ^1A_g^+$      & -49.880 & -49.875 & 0.01 \%  \\ \vspace{0.2cm}
H\"{u}ckel & $E_g$   &  3.233  &  3.286  & 1.61 \%  \\ \vspace{0.2cm}
PPP    &   ${E_g} (1 ^1B_u^-)$ &  3.231  &  3.241  & 0.31 \%  \\  \vspace{0.2cm}
PPP    &   ${E_g} (2 ^1A_g^+)$ &  3.193  &  3.313  & 3.75 \%  \\ \vspace{0.2cm}
PPP    &   ${E_g} (2 ^1B_u^-)$ &  4.229  &  4.346  & 2.77 \% \\ \hline
\end{tabular}
\label{tabular:enr_comp}
\end{center}
\end{table}
{In Fig~\ref{fig: pery-dmrg-scheme}[a], we show the infinite DMRG scheme adopted by us. To improve the accuracy of the calculations, we have also carried out finite DMRG sweeps after constructing the molecule in the infinite scheme. In the finite scheme, the $C_2$ symmetry does not exist when the two blocks of the superblock are of different sizes (see Fig~\ref{fig: pery-dmrg-scheme}[b]). Hence, we apply the $C_2$ operation only when the two blocks are related by this symmetry. We have also constructed the average density matrix using the targeted low-lying states to ensure similar errors in the eigenstates. In Fig~\ref{fig: max-enr} we show the dependence of the PPP ground state energy of perylene in the infinite and finite schemes as the dimension of the block space. In all our studies, we have used a cut-off of at least 800 or more for the block space dimension. To further establish the accuracy for the chosen cut-off we have compared the ground state energy and then exact energy gaps of the low-lying states of perylene in singlet space (Table \ref{tabular:enr_comp}). We see that the ground state energy has an error of 0.02 \% for the H\"{u}ckel model and 0.01 \% for the PPP model. Other quantities such as bond orders as shown in Fig \ref{fig: BO-unsub-s0} and transition dipole moments also show good agreement between the DMRG calculations and exact calculations. The accuracy of the singlet space results gives us confidence to carry out DMRG calculations for the triplet space as well as for the singlet and triplet spaces of donor-acceptor substituted perylene system. The two-photon intensity calculations are done using a DMRG cut-off of 500 in the unsubstituted perylene molecule. The DMRG method was chosen over the exact method as the exact method was computationally prohibitive.}\\
\section{Results and discussions}
{ We have organized this section as follows. We first present the results for unsubstituted perylene and then proceed to give results for the substituted perylenes. Under substituted perylene results, we present the results for the bay substitution and peri substitution separately.}
\subsection{\textbf{Unsubstituted Perylene}}
{ \textbf{Singlet state properties:} In table \ref{enr-trdip}, we present the energies of the singlet states and the transition dipole of the ionic singlet states with the ground state. Since we have used two separate $C_2$ point groups to obtain two states each in the covalent A and ionic B spaces, we only have two unique covalent A levels and three unique ionic B levels.}
\begin{table}[]
\caption{Singlet and Triplet energy gaps (in eV) and transition dipoles to the ground state (in Debye) for the low-lying states of perylene under $C_2(y)$ and $C_2(z)$ symmetries. For the triplet states, we found the same states under both the $C_2$ point groups. For triplets transition dipole moments are from the lowest triplet state to higher lying triplets. Only covalent subspace contains covalent VB diagrams and ionic subspace excludes all covalent VB diagrams.}
\begin{center}
\setlength{\tabcolsep}{4pt} \vspace{0.2cm}
\begin{tabular}{|c|c c c c|}
\hline
                          & State & Energy gaps(eV) & $\mu_x$ & $\mu_y$ \\
                          \hline
\multirow{4}{*}{Singlets} & $2{^\textbf{1}A^{cov}}$       & 3.190      & 0.00                   & 0.00                \\
                          & $^\textbf{1}B_{2}^{ion}$    & 3.232      & 6.67                  & 0.00                 \\
                          & $^\textbf{1}B_{1}^{ion}$    & 4.229      & 0.00                   & 0.00                   \\
                          & $^\textbf{1}B_{3}^{ion}$    & 5.199      & 0.00                   & 6.02                   \\
\hline
\multirow{4}{*}{Triplets} & $1^\textbf{3}B_{1}^{cov}$     & 1.577      & 0.00                     & 0.00                      \\
                          & $1^\textbf{3}B_{2}^{cov}$     & 3.426      & 0.00                     & 0.00                      \\
                          & $^\textbf{3}B_{3}^{ion}$    & 4.717      & 0.00                    & 3.77                    \\
                          & $^\textbf{3}B_{2}^{ion}$    & 4.901      & 6.67                    & 0.00       \\ \hline
                                               
\end{tabular}
\end{center}
\label{enr-trdip}
\end{table}
{ Experimentally, perylene has been widely studied for its electronic states. The lowest singlet absorption is at 2.88 eV in solution phase ~\citep{joblin99,halasinski03}. Our lowest excitation energy for the singlet is at 3.232 eV and the excitation is polarized along the long axis of the molecule \cite{tanizaki78}. Our calculations are for an isolated molecule and assuming a red shift of 0.4 eV in solution,  the lowest singlet energy agrees with the experimental result ~\cite{halasinski03}. We also note that this excitation is polarized along the long axis of the molecule, again in agreement with the experiments. The second lowest energy optical excitation is at 5.199 eV and is polarized along the short axis, which agrees with experiments, if we correct the red shift in solution. The excitation at 4.229 eV is very weak as it has a polarization along the z-axis where the intensity depends on the extent of noncoplanarity of the molecule. Our calculations also predict a two photon state slightly below the one-photon state. This implies that in the gas phase the molecule is not fluorescent, as per Kasha rule. However, in solution, the one-photon state is expected to show a larger red shift and hence in solution perylene is expected to be fluorescent. We have computed the charge gap in perylene, defined as $E_c = E_g(P^+)+E_g(P^-)-2E_g(P)$ , where $E_g(P^+)$, $E_g(P^-)$ and $E_g(P)$ are the ground state energies of cation, anion and neutral perylene molecule. The computed charge gap is 6.358 eV, which makes it a very good insulator with a large exciton binding energy of $\sim$ 3.2 eV.} 

{In principle the two photon absorption cross section can be calculated exactly from the VB theory using a correction vector approach without having to explicitly compute the excited states and the transition dipole matrix elements \citep{soossr89,sr-smmat}. However, numerically we found this approach to be very CPU intensive. Instead we have calculated the two photon transition matrix elements within the DMRG scheme, using a correction vector approach \cite{jha04}. The first order correction vector $\phi_i^1 (\omega)$ for TPA calculation involves solving the linear system 
\begin{equation}
(H-E_G\pm\hbar\omega_n) \left| \phi_i^{(1)} (\pm\omega_n) \right\rangle = \widetilde{\mu_i}\left|G\right\rangle
\end{equation}
where $\widetilde{\mu_i}$ is the dipole displacement operator for the $i^{th}$ component $(i = x, y, z)$, $E_G$ is the ground state energy, $\hbar\omega_n = E_{nA_{g}}/2$ corresponds to the two photon resonance frequency and H is the Hamiltonian in the covalent A subspace. We have employed the $C_2$ symmetry about the z-axis, perpendicular to the plane of the molecule. In the DMRG procedure we have used a cut-off of 500 block states and solved for $\phi_i^{(1)}(\pm\omega_n)$ by a small matrix technique \cite{sr-smmat}.
The TPA transition matrix element $S_{ij}(\omega_n)$ is given by
\begin{equation}
S_{ij}(\omega_n) = \left[\left\langle\phi_i^{(1)}(-\omega_n)\left|\widetilde{\mu_j} \right| nA\right\rangle + \left\langle \phi_j^{(1)}(-\omega_n)\left|\widetilde{\mu_i}\right| nA \right\rangle \right]
\end{equation}  
The orientationally averaged two photon cross section can be computed from $S_{ij}(\omega_n)$ as
\begin{equation}
\delta_{TP} = \dfrac{1}{30} \sum_{ij} \left( 2S_{ii}S_{jj}^* + 4S_{ij}S_{ij}^* \right)
\end{equation}
 The TPA cross section is given by
 \begin{equation}
 \sigma_{G \rightarrow nA} = \dfrac{4{\pi^3}{a_0^5}{\alpha}{\omega_n^2}g(\omega_n)}{c\Gamma} \delta_{TP} 
 \end{equation}
 where $a_0$ is the Bohr radius, $\alpha$ is the fine structure constant and c is the velocity of light. We have assumed a brodening parameter $\Gamma$ of 0.1 eV and $g(\omega_n)$ to be a constant set to unity. In Table \ref{table: TPA} we present the results for the TPA cross section for the 3 low-lying states in the A space.
\begin{table}
\caption{TPA cross-section for perylene, obtained from DMRG calculation for the lowest three two photon states. Here $\hbar\omega_n$ is half of the energy gap from ground state (in eV). Transition matrix elements and average TPA cross section are given in atomic units and TPA cross-section is in Goeppert-Mayer units. }
\begin{tabular}{|c|c|c|c|c|c|c|}
\hline
 State & $\hbar\omega_n = E(nA)/2$  & $S_{11}$ (in a.u.)  &  $S_{12}$ (in a.u.)  &  $S_{22}$ (in a.u.) & $\delta_{TP}$ (in a.u.) & $\sigma_{nA} (GM)$ \\ \hline
2A & 1.62 &  0.04 & -0.67 & -0.27  & 358.40 & 4.47 \\ \hline
3A & 1.67 &  0.04 &  0.70 &  0.28 &  383.26 & 5.05 \\ \hline
4A & 2.12 &  0.05 &  1.20 &  0.45 & 1116.24 & 23.63\\ \hline
\end{tabular}
\label{table: TPA}
\end{table}

{We find that the intense TPA absorption is to the $4A_g^{cov}$ state, while the first two TPA states have similar but smaller intensity than the $4A_g^{cov}$ absorption by almost a factor of five.}}

\begin{figure}[]
\begin{center}
\includegraphics[width=0.6\textwidth]{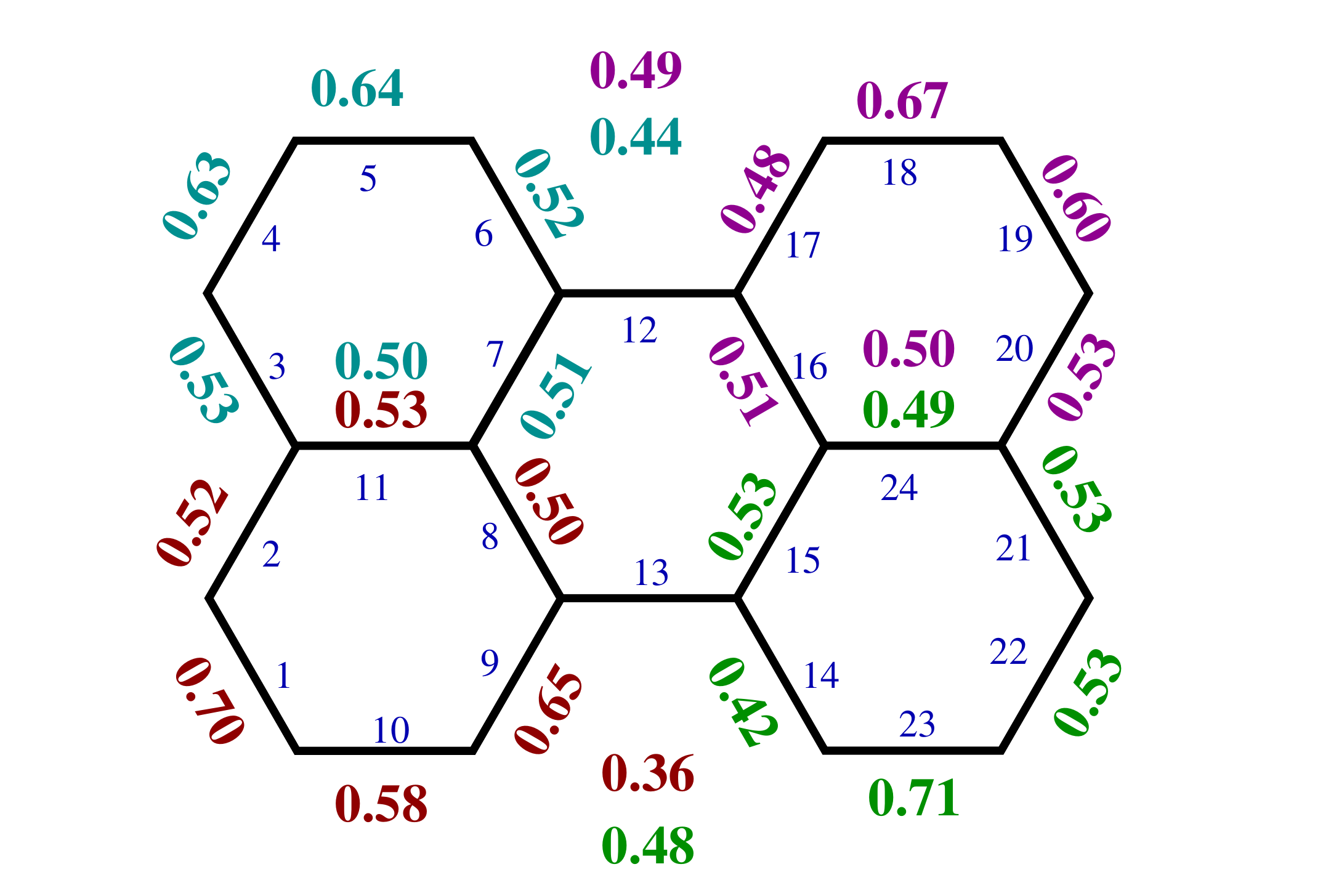}
\end{center}
\caption{Unique Bond orders for unsubstituted perylene. Inside the perylene bond indices are shown in blue color. Bond orders for lowest singlet state (in red), dipole-allowed singlet state (in cyan), and two photon state (in green) are shown. Triplet lowest state bond orders are shown in magenta color.}
\label{unsub-BO-s1}
\end{figure}

{Our calculations are for an idealized geometry. The equilibrium geometry will be distorted from the idealized picture. The bond orders $p_{ij}$ which are defined as $-\dfrac{1}{2} \left\langle \psi \left| \hat{E}_{ij}+\hat{E}_{ji} \right| \psi \right\rangle$, for an eigenstate $\psi$ give an insight into the distortion that the molecule has in the equilibrium geometry of the state $\psi$. Schematic picture of the equilibrium geometries, based on the bond order for the ground state, the one photon and two photon states are shown in Fig \ref{unsub-BO-s1}.}

{ The ground state geometry can be described as two weakly linked naphthalene units \cite{dickens11}. Within the naphthalene units, the bonds not linked to the sites common to the two rings, has a structure similar to butadiene. In the one photon state, the two naphthalene units are more strongly linked and the central benzene ring is more uniform. The bonds between carbon atoms with three coordinates (site nos. 7, 14 and 9, 12 ) become stronger and hence are expected to shrink. The two photon state shows an altogehter different geometry. The horizontal bonds in the benzene rings at the corners are shorter or stronger bonds, the central ring has the same bond order pattern as in the one-photon state. The bond between the bay site and the central ring becomes weaker and stretched. The bond order pattern suggests that the stokes shift in the fluorescence of the molecule should be significant.}\\

{\noindent \textbf{Triplet state properties:} The triplet gap in perylene is at 1.577 eV, very close to the experimental value of 1.534 eV \cite{clarke69,offen67,h61} found in the crystal. The lowest triplet is a covalent triplet and hence only a weak red shift from gas phase to crystalline phase is to be expected. The second lowest triplet in the covalent space is at 3.426 eV. The ionic triplets are at 4.717 eV and 4.901 eV. Although these two states are close in energy the first ionic state has x-polarization while the second has y-polarization in a one-photon triplet-triplet absorption spectra. The energetic of the lowest excited singlet at 3.232 eV and the two lowest triplet states at 1.577 eV and 3.426 eV satisfy the energy criterion for singlet fission. The triplet excitons of resonably high energy and the second lowest triplet exciton is above the singlet state, preventing an intersystem crossing to this state.} \\

{\noindent \textbf{Spin densities:} The spin density of the lowest triplet state is mainly localized at the peri and bay sites of the molecule. The central ring sites has almost no spin density while the remaining sites have slight negative spin densities. The spin densities of the $2^{nd}$ covalent triplet state is large at the bay sites but smeared out over rest of the molecule. The spin densities of the lowest ionic triplet state have similar distribution as in the lowest covalent triplet. In the second lowest ionic triplet state, the peri sites have significantly higher spin density, in contrast to the spin density distribution in the second covalent triplet exciton.}\\

{\noindent \textbf{Bond orders:} Regarding the bond orders of these states, the bonds connecting the two naphthalene units are weak in all the cases, although in the lowest covalent and ionic triplets, it is larger than in the second covalent and ionic triplets. In all the states, the bonds with at least one triply coordinated carbon site are slightly stronger than these inter-naphthalene bonds. The bonds involving the bay and peri sites with the corner sites are stronger than the other bonds. In the first covalent and second ionic triplets the corner and bay bond is weaker than the corner and the peri bond while this switches for the second covalent and the first ionic triplet states.}\\
\begin{figure}[]
\begin{center}
\includegraphics[width=0.6\textwidth]{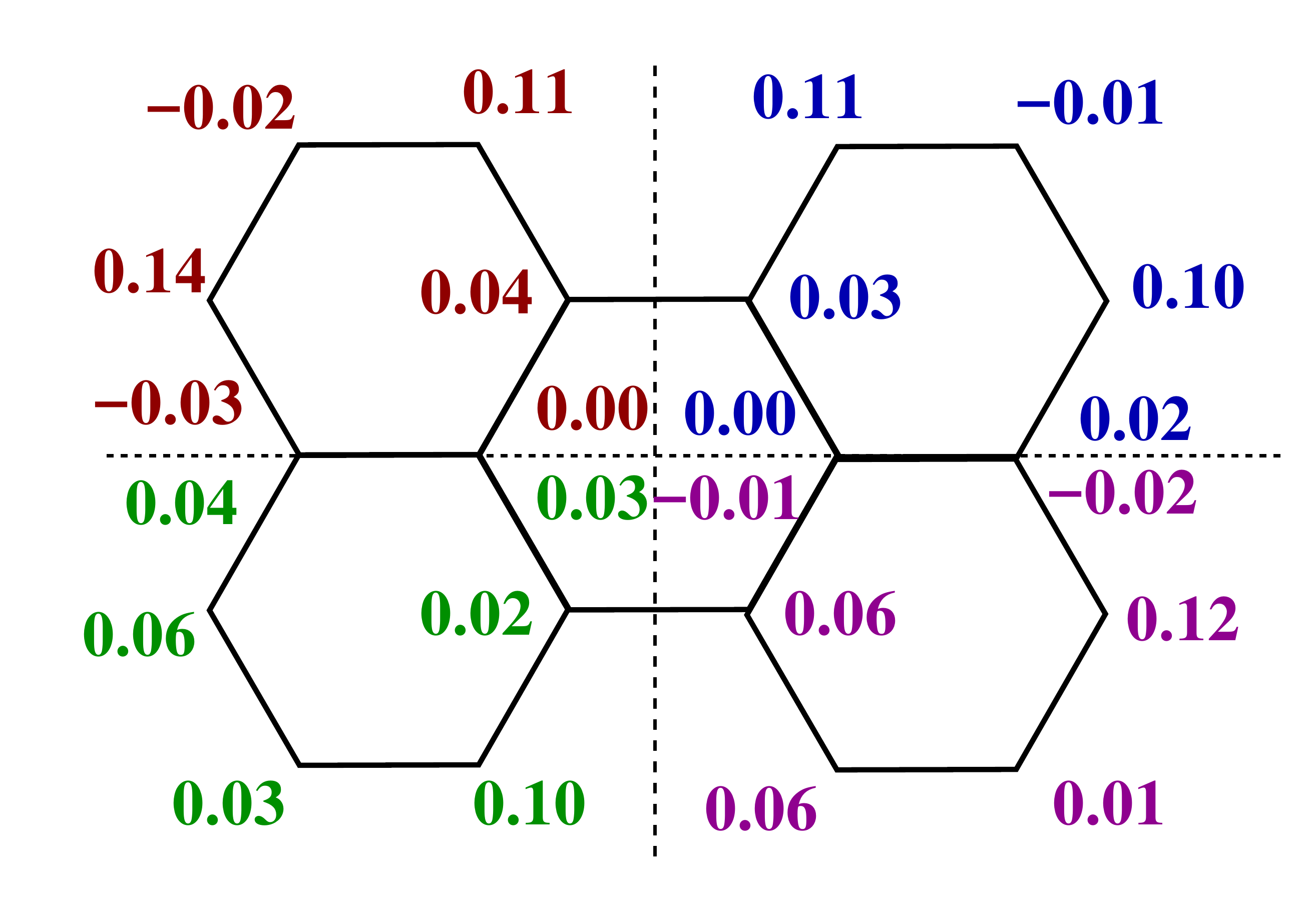}%
\end{center}
\caption{Unique Spin densities in unsubstituted perylene for lowest triplet state (top left, in red) and in second covalent triplet state (bottom left, in green); first ionic triplet state (top right, in blue) and second ionic triplet state (bottom right, in magenta).}
\label{fig: unsub-spnden}
\end{figure}
\subsection{Substituted Perylene}
\begin{figure}[]
\begin{center}
\includegraphics[scale=0.4]{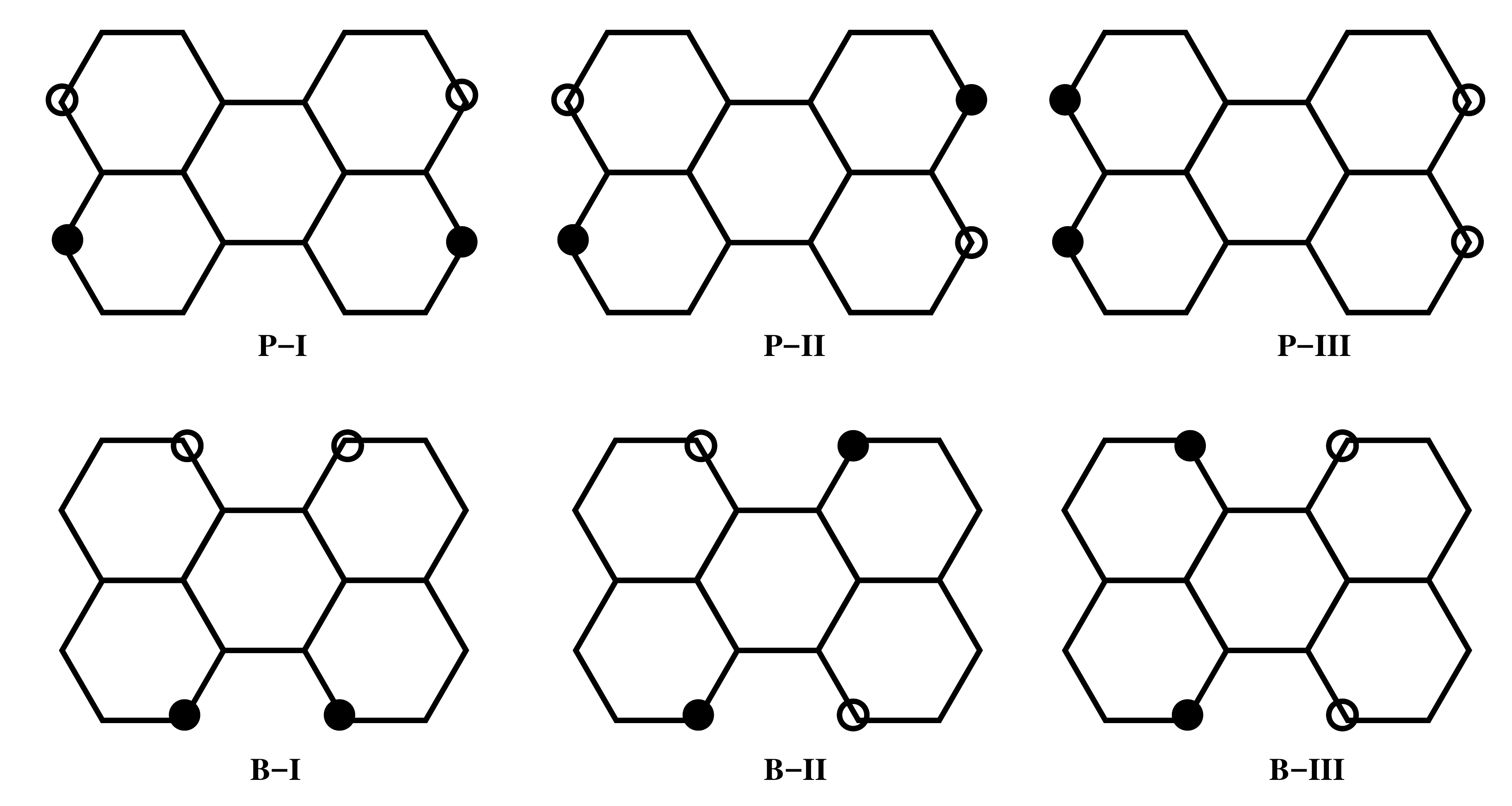}
\setlength{\belowcaptionskip}{0.5pt} 
\end{center}
\caption{Structure of substituted perylene molecule where filled circles represent donors and open circles represent acceptors. P-I, P-II and P-III correspond to three arrangements of subsitution at the peri site nos. 2, 4, 17, 19, and B-I, B-II and B-III correspond to the three arrangments of substitution at the bay site nos. 6, 10, 11, 15.}
\label{fig: sub-scheme}
\end{figure}

\begin{figure}[]
\begin{center}
\subcaptionbox{P-I}{\includegraphics[width=0.30\textwidth]{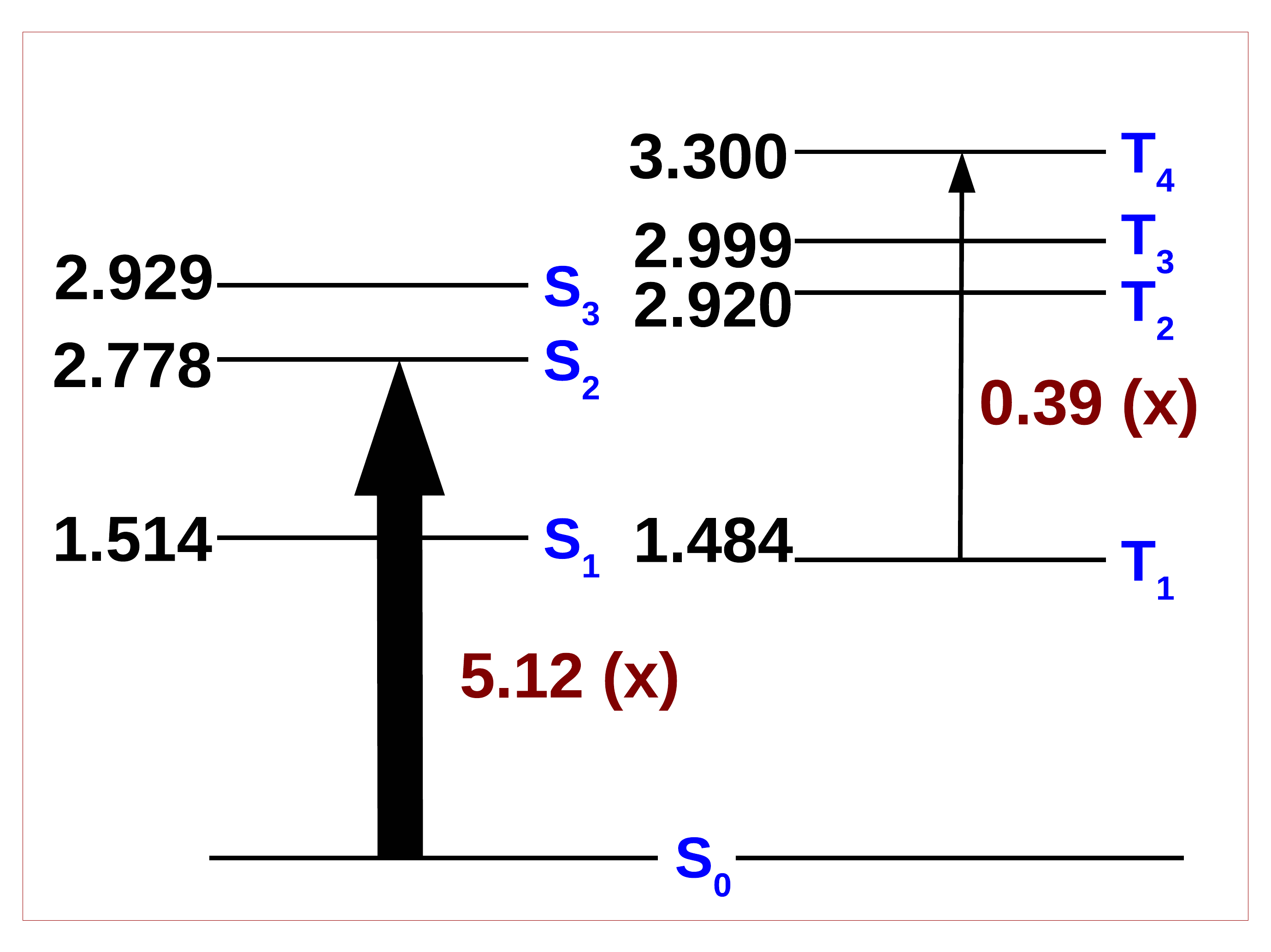}}%
\hspace{0.1cm}
\subcaptionbox{P-II}{\includegraphics[width=0.30\textwidth]{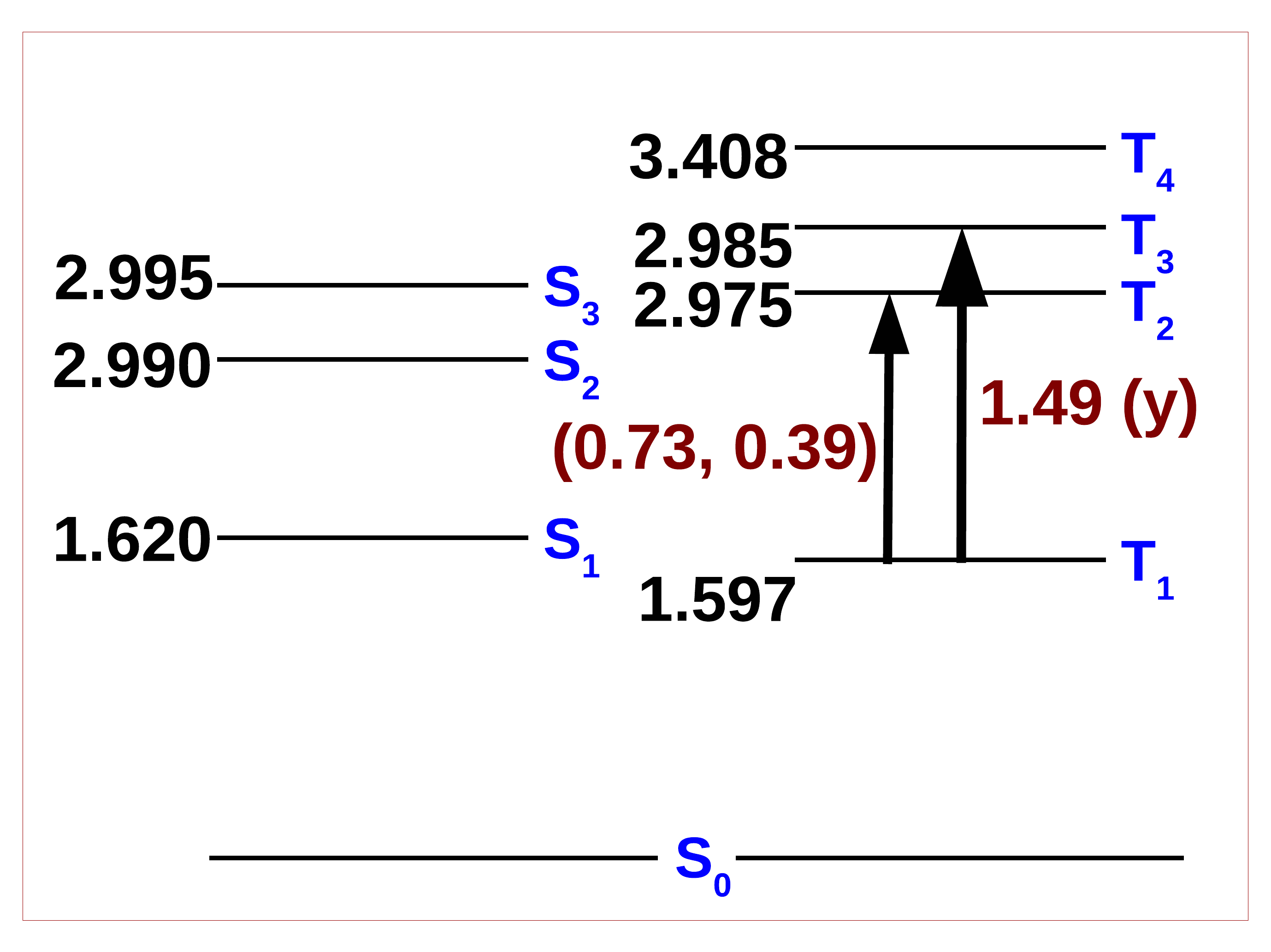}}%
\hspace{0.1cm}
\subcaptionbox{P-III}{\includegraphics[width=0.30\textwidth]{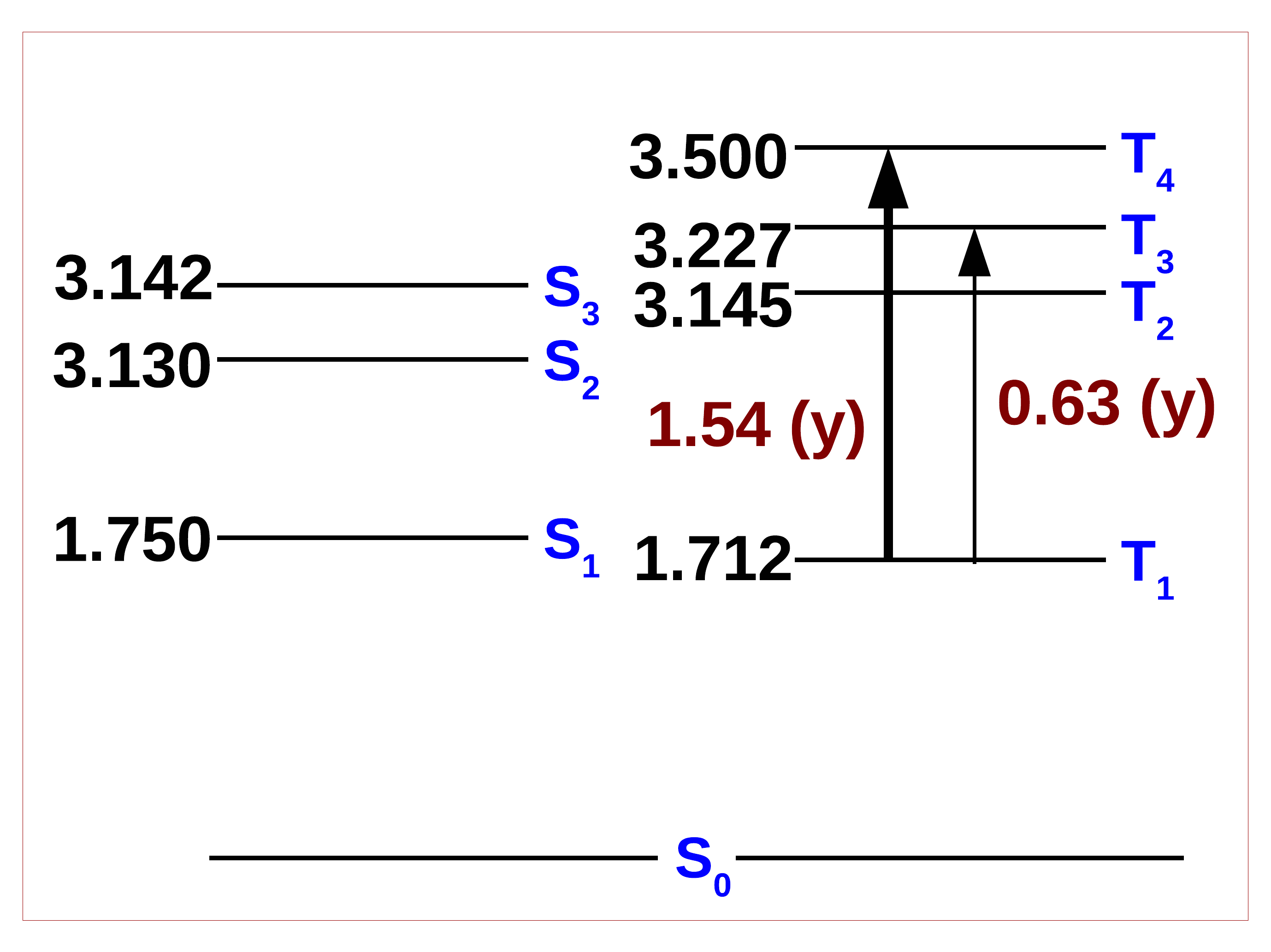}}%
\vspace{0.2cm}
\subcaptionbox{B-I}{\includegraphics[width=0.30\textwidth]{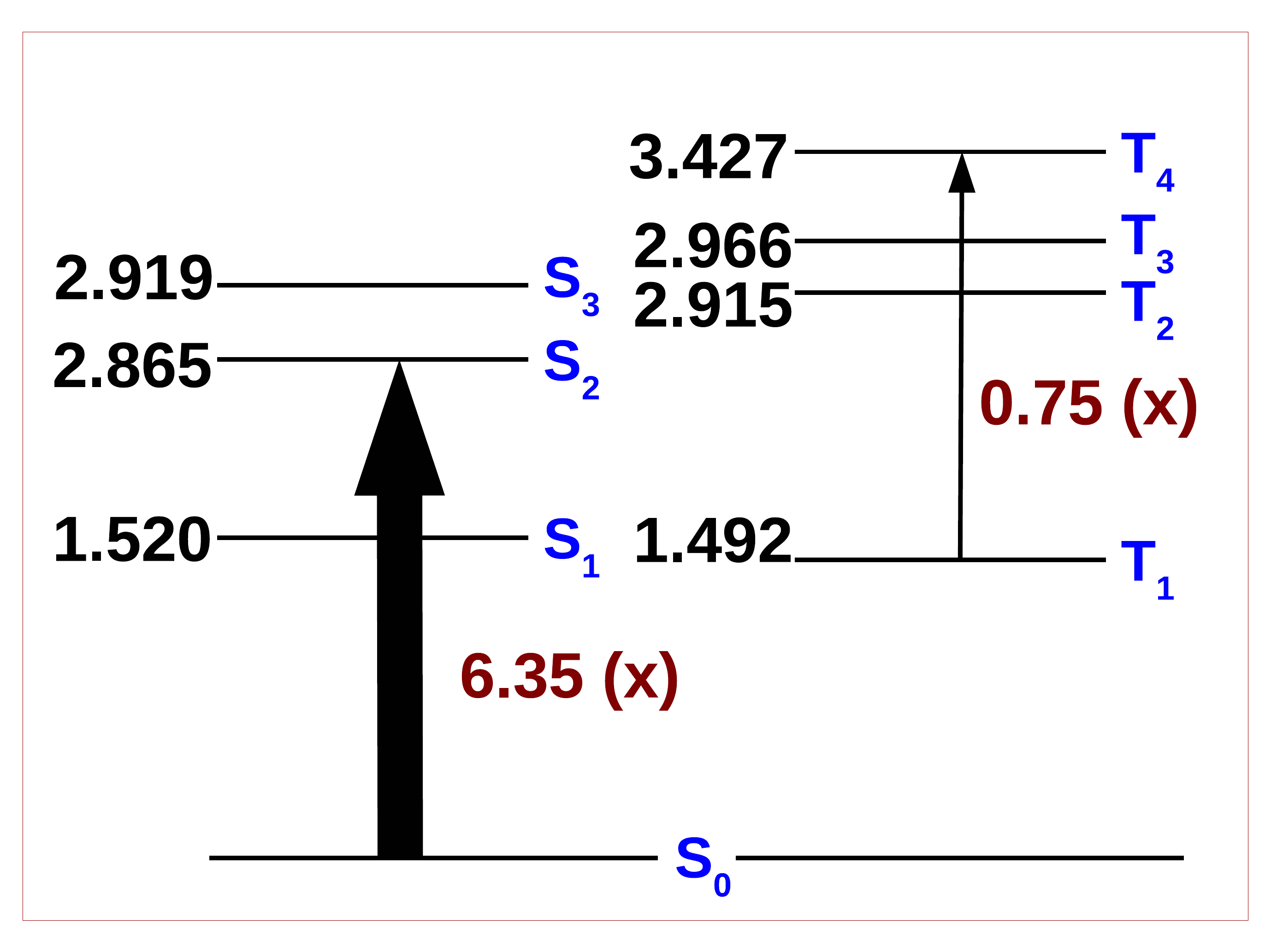}}%
\hspace{0.1cm}
\subcaptionbox{B-II}{\includegraphics[width=0.30\textwidth]{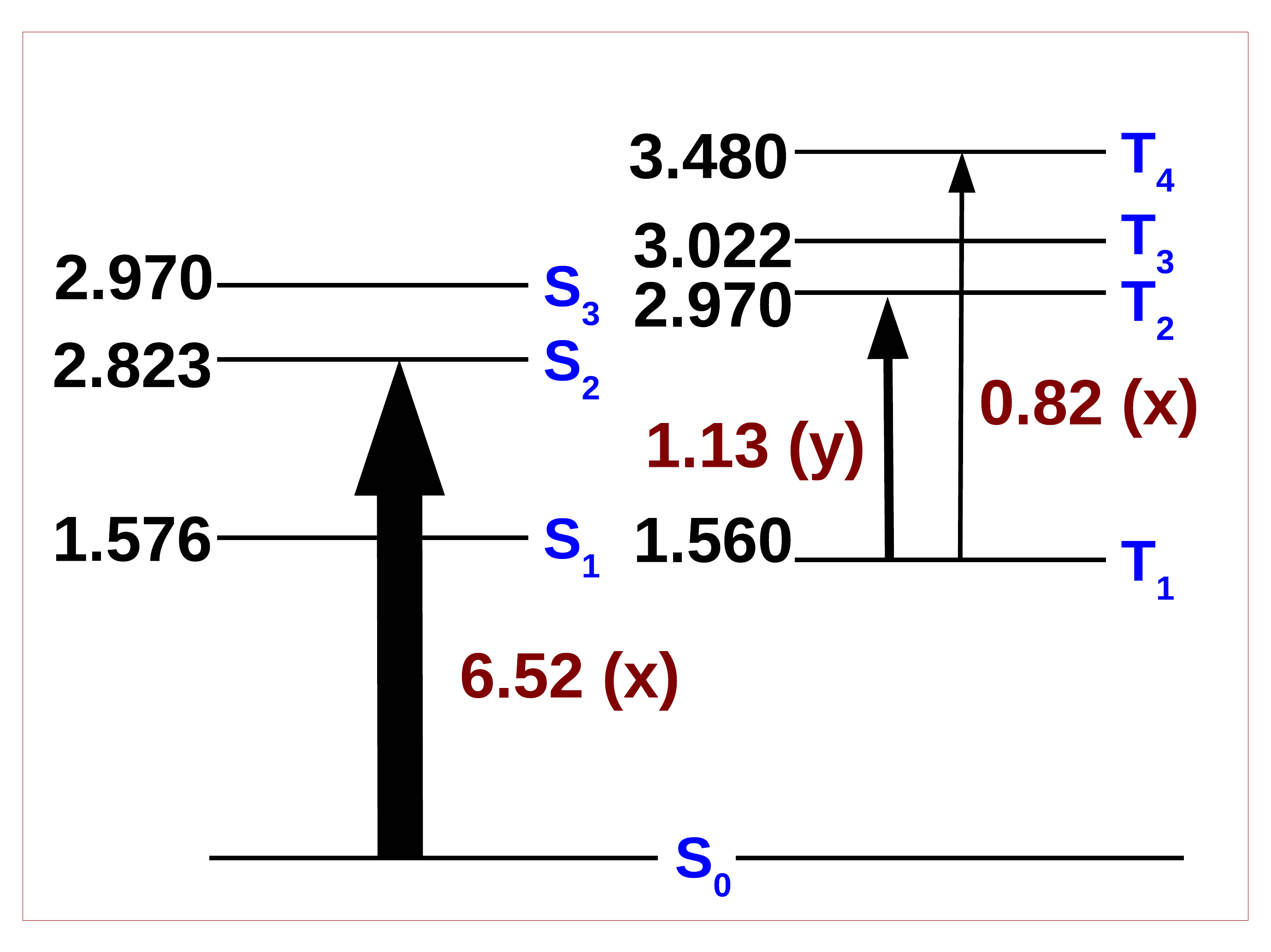}}%
\hspace{0.1cm}
\subcaptionbox{B-III}{\includegraphics[width=0.30\textwidth]{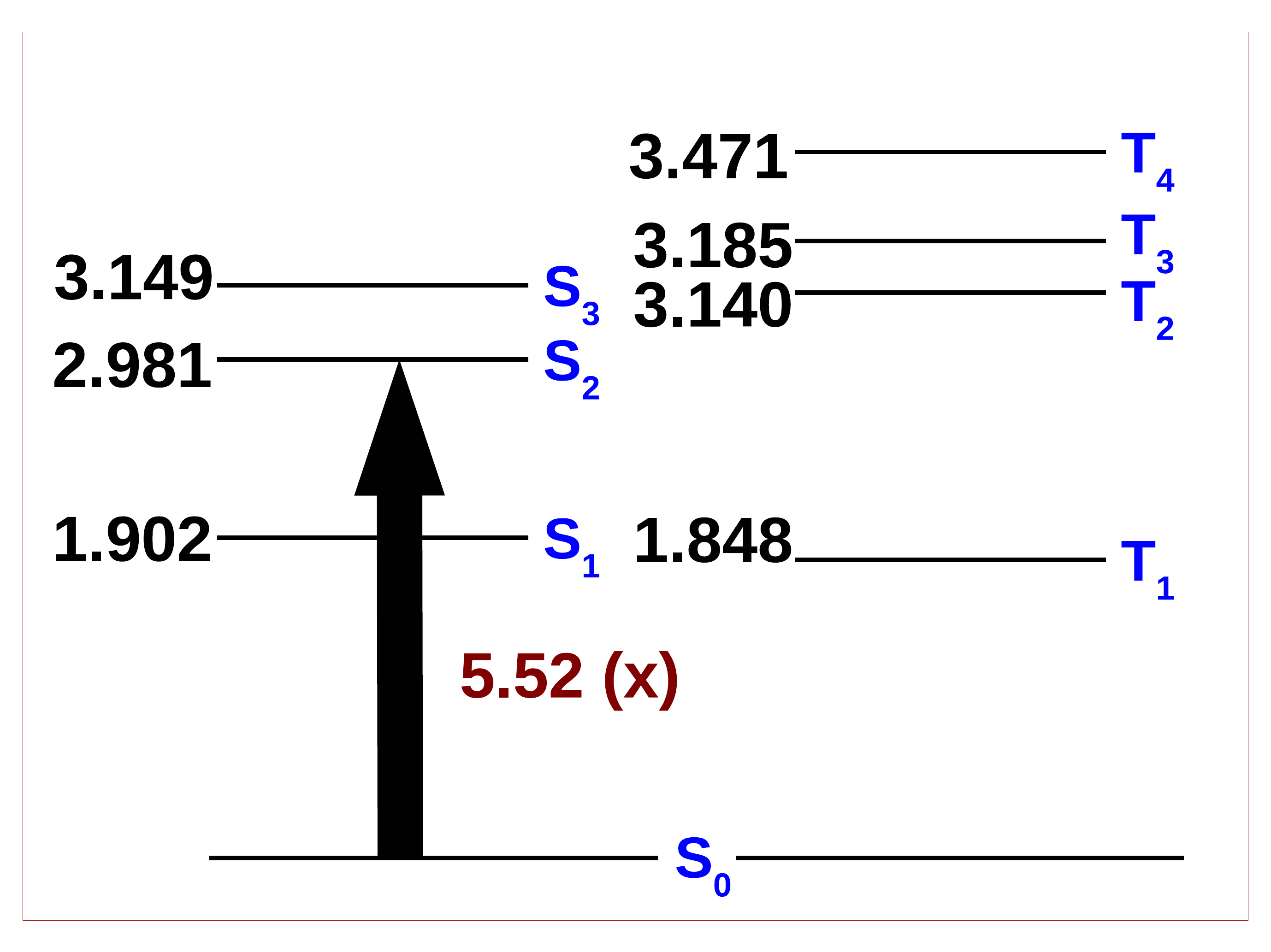}}%
\end{center}
\caption{Transition dipole moments for vertical transitions from ground state to dipole allowed states in singlet as well as triplet states for $\epsilon=2 eV$ in all six arrangments of peri and bay substitution. Inside each panel, line at the bottom shows ground state of corresponding perylene system and lines lying above (in left) show singlet excited states and (in right) triplet excited states (in blue color). The energy gaps from ground states are in black color. The transition dipole allowed levels are given in red color (also connected by arrows) and the letter in bracket gives the polarization direction of the transition dipole.}
\label{fig: sub-trdip}
\end{figure}

{In substituted perylenes, two types of substitutions have been widely studied. They are the peri substitution and the bay substitutions. In the peri substituted systems, the diimides and the carboxilates \citep{chen14,huang11,qian08,gisslen09} are the substituents that have been studied, while in the bay substitutions, the substituting groups are poly thiophene and bromide groups \citep{B901847F,GANESAMOORTHY201899,schill18,li17,voll17}. In our studies, we have considered symmetric substitutions by 2 donors and 2 acceptors with $C_2$ symmetry about one of the axes, with all the substituents only at the 4 peri sites or only at the 4 bay sites. The three D-A substituted systems we have studied in the two (peri, bay) cases are shown in Fig \ref{fig: sub-scheme} . We have not considered any ortho substitutions or mixed peri and bay substitutions. We have assumed equal strength for the D and A groups by assuming $\left|{\epsilon_D}\right| = \left|{\epsilon_A}\right|$ where $\epsilon$ are the site energies; negative $\epsilon$ corresponds to an acceptor site while a positive $\epsilon$ corresponds to a donor.}\\
{The substitution strength we have studied correspond to $\left|{\epsilon}\right| =$ 1, 2 and 3 eV. For the sake of brevity we present the results for $\left|{\epsilon}\right| =$ 2 eV. The remaining results are presented in supporting/ supplementary information. 
In Fig~\ref{fig: sub-trdip} we summarize the energies and transition dipoles of a few low-lying singlet and triplet states. In the singlet space we note that there is a singlet state in the energy range 1.5 to 1.9 eV, while the lowest excited singlet energy reported in the unsubstituted perylene is at 3.19 eV. This is because in the unsubstituted case, the only covalent singlet subspace we explored was the A space. This singlet state of the substituted perylene is therefore arising from the covalent space of $B_x$ symmetry of the unsubstituted perylene. This is also borne by the fact that the transition dipole from the ground state to this state is less than $10^{-2}$ Debye. Indeed, this also holds true of the $S_3$ state in all the substituted molecules. The excitation to $S_2$ is allowed in P-I, B-I, B-II and B-III but is again almost forbidden in P-II and P-III. The dipole allowed excitation is x-polarized in P-I, B-I, B-II and B-III and all have very similar intensities, besides having roughly the same energy gap of 2.826 $\pm$ 0.048 eV. This value is about 0.35 to 0.45 eV red shifted from that of the parent molecule. The sign and magnitude of the ground state dipole moment of all these systems is also in line with the substitution pattern.}

{The lowest triplet state in all these systems lies between 1.484 eV and 1.712 eV while in the parent molecule the lowest triplet lies at 1.577 eV. The charge gaps in substituted perylenes are also large and show small variation across substitution. For $\left|\epsilon\right| = 2 eV$, the charge gaps are 5.98 eV, 6.22 eV, and 6.19 eV for P-I, P-II, and P-III and 6.03 eV, 6.06 eV, and 6.17 eV for B-I, B-II and B-III. Hence, the substitutions hardly affect the electrical transport in the perylene system.}\\

\begin{figure}[]
\begin{center}
\subcaptionbox{Bond orders}{\includegraphics[width=0.49\textwidth]{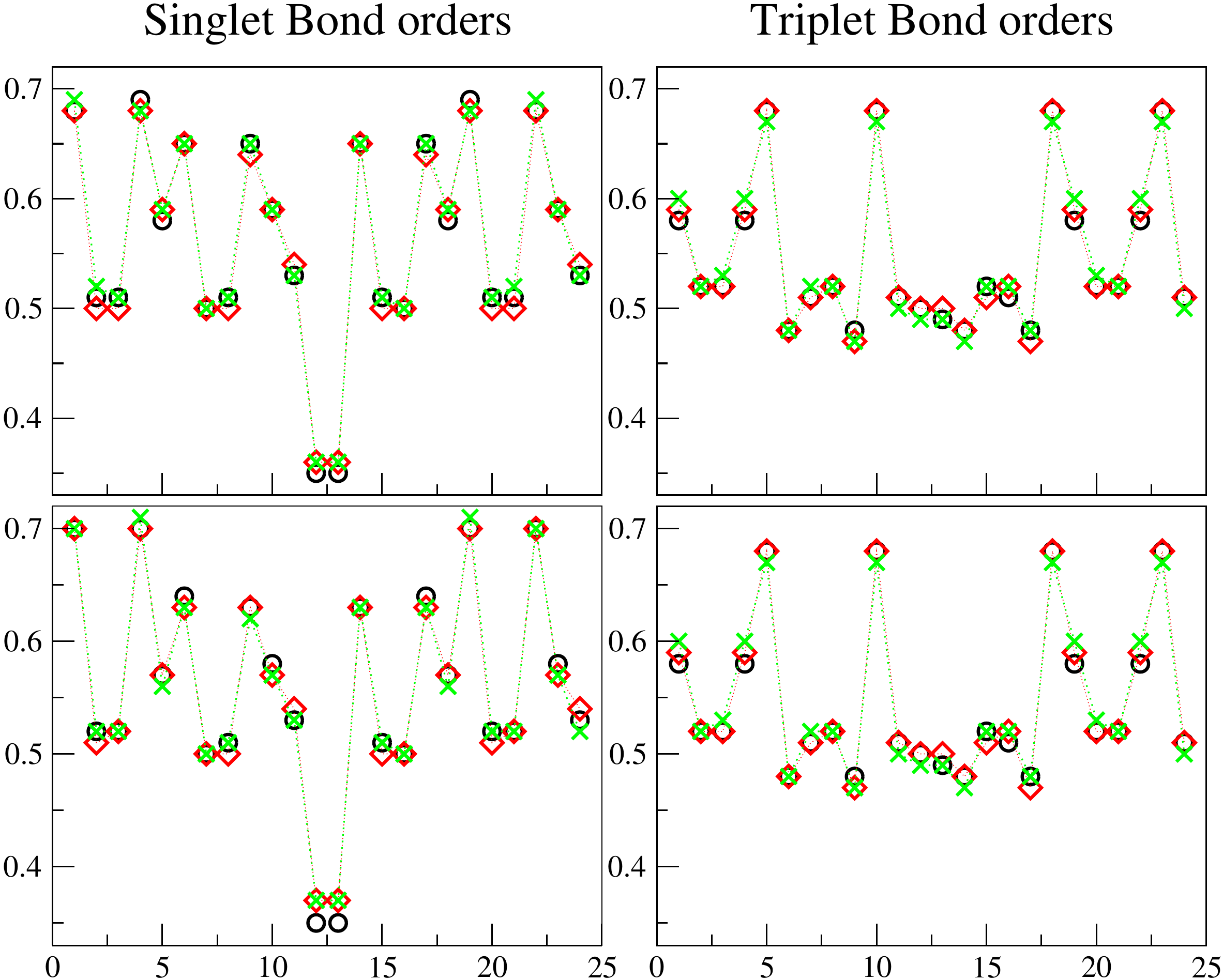}}%
\hspace{0.1cm}
\subcaptionbox{Charge densities}{\includegraphics[width=0.49\textwidth]{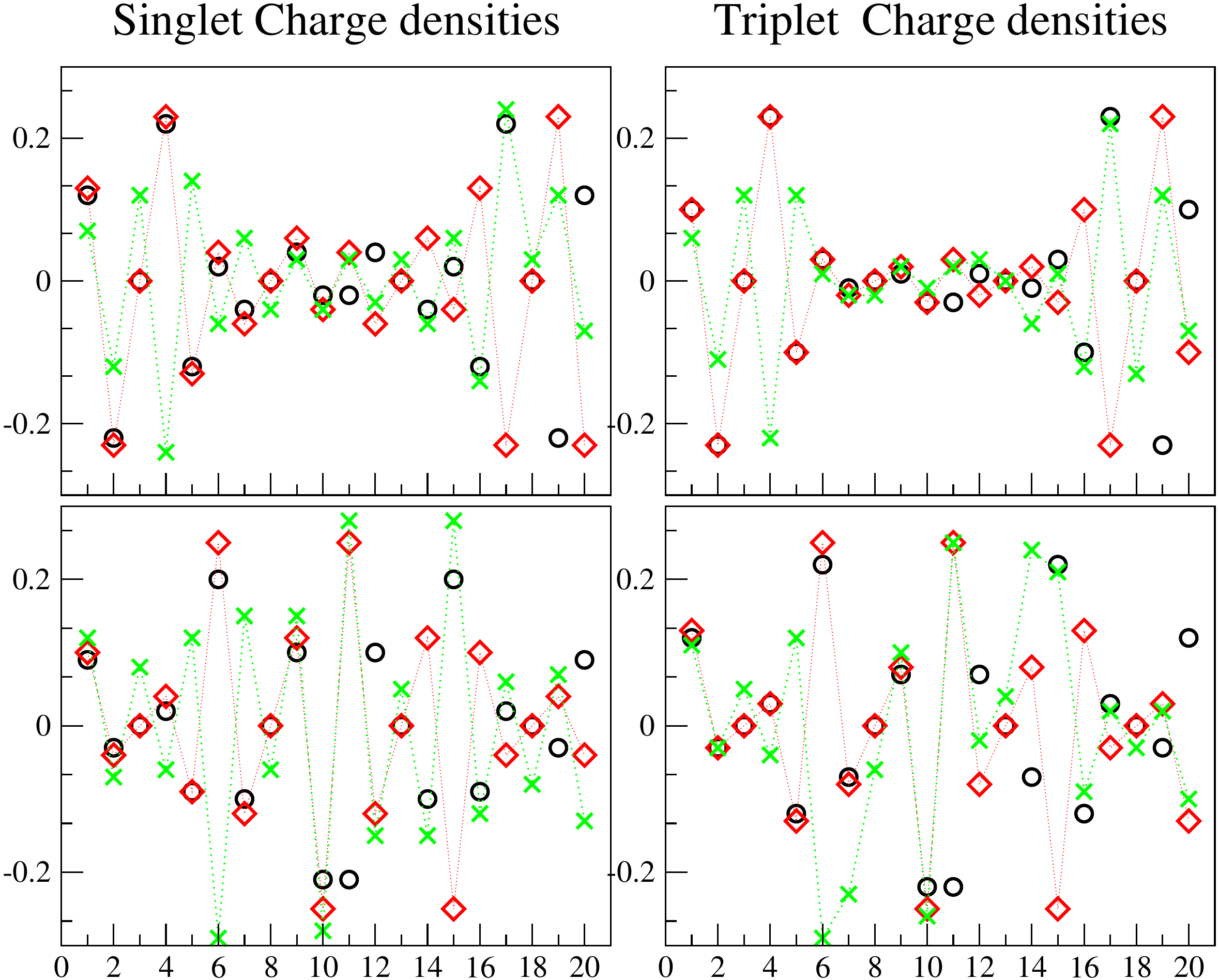}}%
\end{center}
\caption{Bond orders, charge densities and spin densiies of triplets for substituted perylenes with D/A strength of $\epsilon= 2 eV$. Top panels are for peri substitution and bottom panels are for bay substitutions. The color coding and symbols are the same as in Fig 9.}
\label{fig: sub-BO}
\end{figure}

{\noindent \textbf{Bond orders in substituted perylene:} We have studied the bond orders in the singlet and triplet states of these systems and they are summarized in Fig~\ref{fig: sub-BO}. In both the singlet states and the triplet states the bond orders depend only weakly on the substitution sites (bay or peri) and also on the specific arrangement of the donor and acceptor groups. In fact the D-A substitution does not affect the bond orders significantly. The ground state bond orders are consistent with two weakly coupled naphthalene units. In the triplet state, the strongest bonds are between the ortho and the bay sites. The bonds between ortho and the peri sites are slightly weaker. The bonds between the two naphthalene units strengthen and show approximately the same bond order as the remaining bonds.} \\

{\noindent \textbf{Charge and spin densities:} The charge densities in the substituted compounds show the expected trend (see Fig~\ref{fig: sub-BO}) a large positive charge at the donor sites and a large negative charge at the acceptor end with the magnitude at other sites approching zero as the distance from the D-A sites increase. The main difference between the singlet states and the triplet states is that away from the substituted sites, the sites are more neutral in the later case.}\\

\begin{figure}[]
\begin{center}
\includegraphics[width=0.65\textwidth]{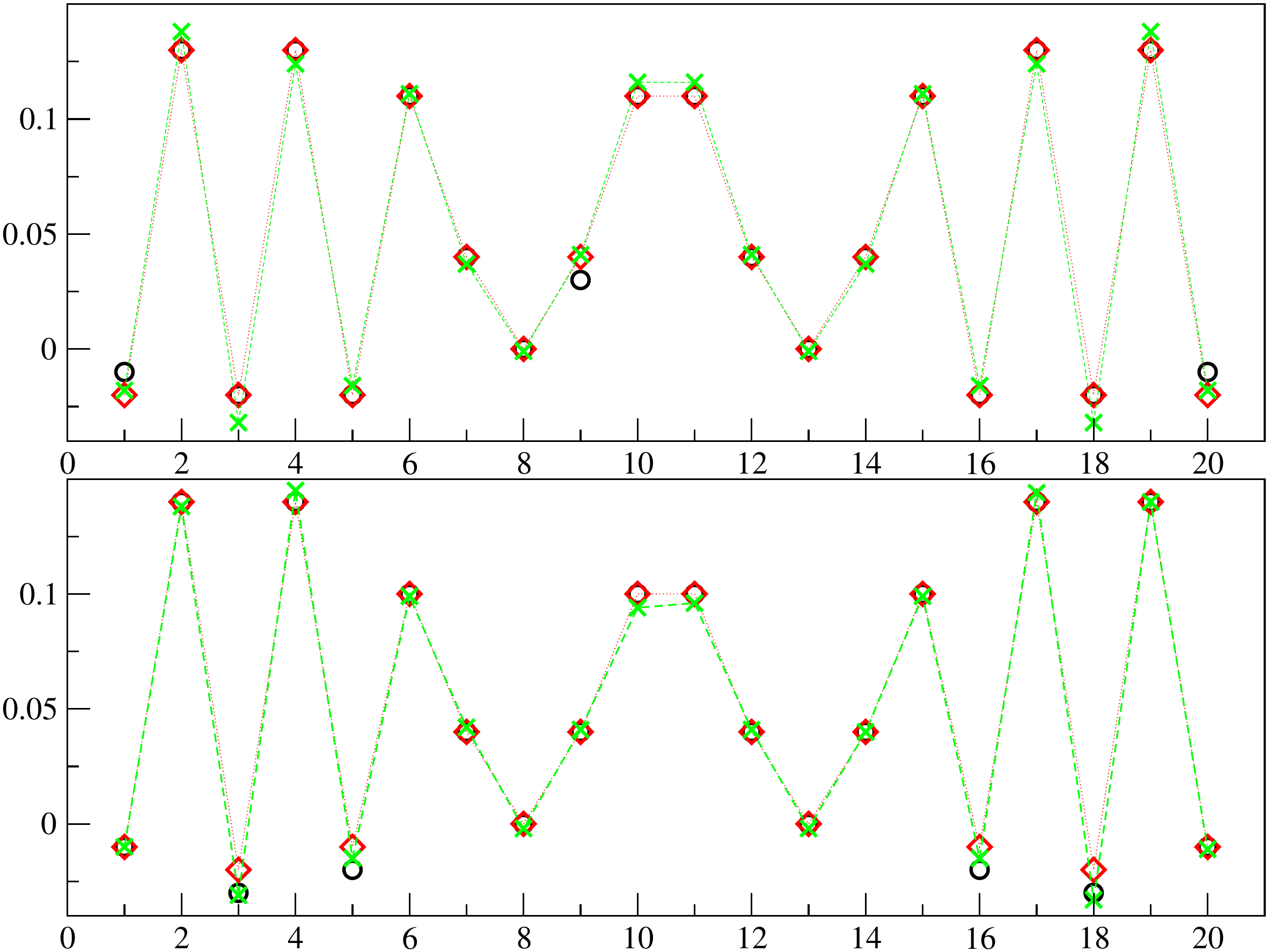}
\end{center}
\caption{Spin densities in lowest triplet state in all the substitution arrangements of bay and peri substitutions for $\epsilon=2 eV$ are given. The top panel correspond to peri substitution and bottom panel to bay substitutions. Here black circles show first arrangement of D-A substitution pattern I, red diamond is for the second substitution pattern and green crosses is for the third substitution pattern.}
\label{fig: sub-spnden}
\end{figure}

{The spin densitites in the lowest triplet states of these molecules show very similar trends independent of the substitution pattern or sites of substitutions (see Fig~\ref{fig: sub-spnden}). The spin density is large and positive at the peri sites. The bay sites have slightly smaller positive spin densities. The sites which connect the two naphthalene units have small positive spin densities. }

{The remaining sites in the central ring have small negative spin density, while the sites 3 and 18 (see Fig~\ref{fig: sub-spnden}) have the largest negative spin densities. The positive spin densities are mainly on the peri and the bay sites, while the ortho sites have negative spin densities. This is consistent with the bond orders where the bonding between ortho and bay and ortho and peri are the strongest in the triplet states.}\\
\section{Summary} {We have presented the energies and other properties of the low-lying electronic states of perylene within a Parisar-Parr-Pople model. We have obtained model exact results in the singlet subspace which spans a 5.92 billion Hilbert space. We have employed various $C_2$ symmetries and electron-hole symmetry to factorize the Hilbert space to 1.48 billion states in the subspace of the ground state and 1.48 billion states in the dipole allowed B subspace. We obtain two one-photon states with energies 3.232 eV and 5.199 eV with transition dipoles of 6.67 D and 6.02 D along the x and y axis respectively. The exact triplets of perylene are not accessible to us in the VB approach. We have used a DMRG approach to access the triplet states with high accuracy. We find the spin gap to be about 1.58 eV. The triplet-triplet absorption are at 2.14 eV and 3.32 eV with transition dipoles of 3.77 D and 6.67 D along the y and x axis respectively. The bond orders show that the molecule can be described as two weakly fused naphthalene molecules in the singlet ground state and slightly strongly fused naphthalene units in the one-photon state. The triplet spin densities are mostly confined to the peri and bay sites. The two photon absorption cross-section from DMRG calculations yields  a strong TPA absorption at 2.237 eV and weaker TPA absorptions at 1.59 eV and 1.73 eV. In perylene, the location of the lowest excited singlet and the lowest triplet states are well suited for singlet fission.\\

D-A substituted perylenes were studied with 2D and 2A groups substitutions mimicked by using equal site energies of opposite signs for the D and A groups. The D and A substitutions are made at all the bay sites or all the peri sites. All the three types of bay substitutions and a single type of peri substitution lead to lowering the one-photon gap to between 2.78 eV and 2.98 eV. The spin gap gap varies between 1.48 eV and 1.85 eV. While the shift in singlet and triplet energies does not favor singlet fission, the substitution is helpful in tuning the band gap for light emitting devices. The bond orders as well as the spin densities in the triplet states are similar to those found in unsubstituted perylenes.}\\ 

\noindent
\textbf{Associated content:}\\
{\textbf{Supporting information:} We have given the data for all three arrangements of substitutions with site energy $\epsilon = 1.0 eV$ and $\epsilon = 3.0 eV$ in this section, \\
\noindent
i)  Figure for energy gap trend against intensity of substitution in both the cases of substitution.
 
\noindent
ii) Energy gaps and associated transition dipole moments table for various states in singlet and triplet states of peri and bay substituted perylenes. 

\noindent
iii) Bond orders in lowest states for singlet and triplet space in peri and bay substituted perylenes. 

\noindent
iv) Charge densities in lowest states for singlet and triplet space in peri and bay substituted perylenes. 

\noindent
v)  Spin densities in triplet lowest states in peri and bay substituted perylenes. }\\

{\noindent
\textbf{Corresponding Author} E-mail: geetgiri1@gmail.com \\
\textbf{Notes} The authors declare no competing financial interest.\\
\textbf{Acknowledgement} We thank the Department of Science and Technology, India, for financial assitance through various projects and Thematic Unit of Excellence in Computational Material Science for computer resources.\\

\end{document}


\begin{figure}[]
\centering
\includegraphics[scale=0.50]{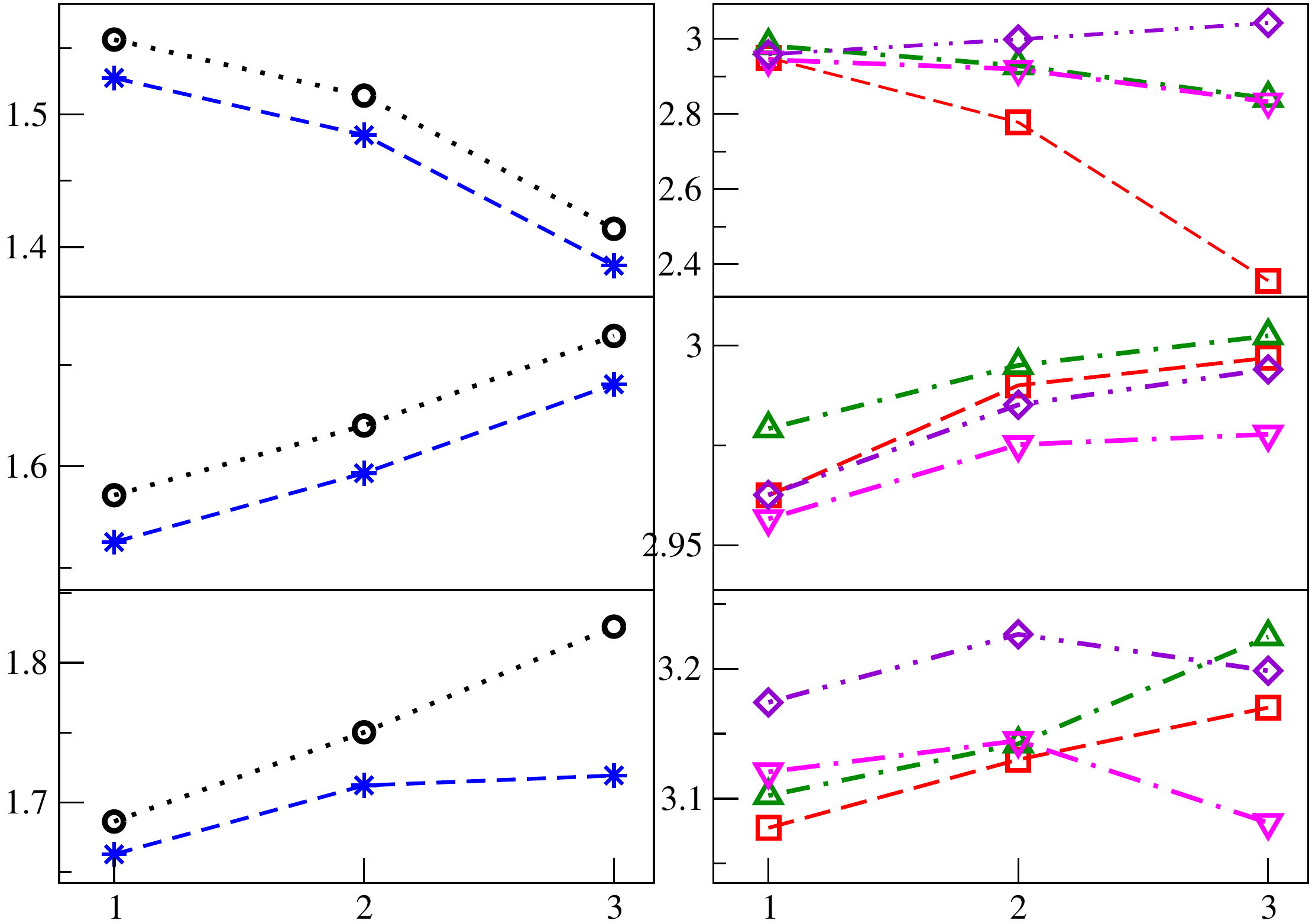}
\setlength{\belowcaptionskip}{-5pt} 
\caption{In three panels three arrangements of peri-substitution are given. Refer figure 9 in paper for P-I, P-II and P-III.
In left panels black circles show lowest singlet states and blue stars are lowest triplet states corresponding to substitution strength $\epsilon = 1, 2, 3$. In right panels red squares and green upper traingles represent first and second excited singlet states whereas magenta lower traingles and violet diamonds symbolize the first and second excited triplet states. }
\label{fig: sub-peri}
\end{figure}
\begin{figure}[]
\centering
\includegraphics[scale=0.50]{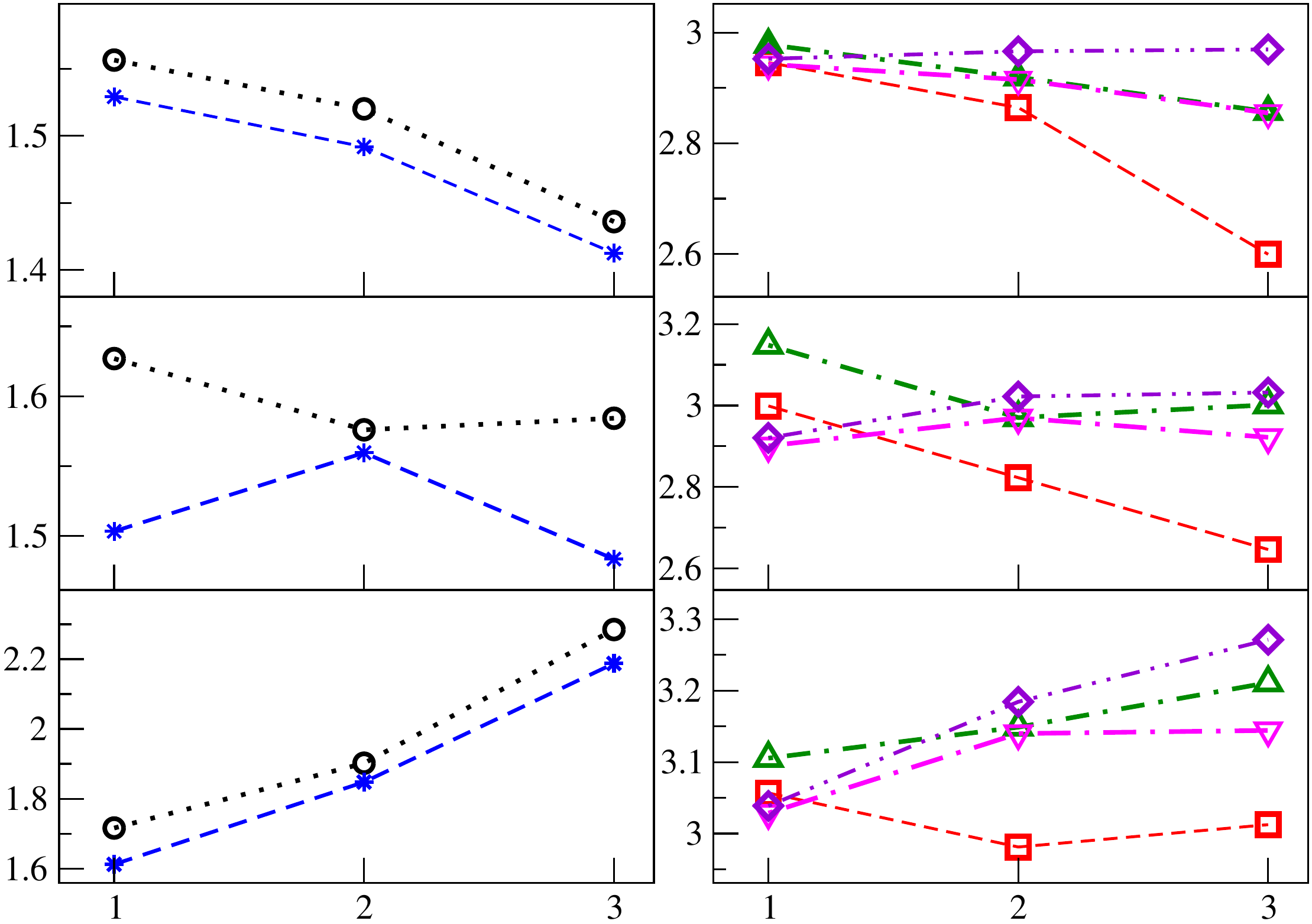}
\setlength{\belowcaptionskip}{-5pt} 
\caption{In three panels three arrangements of bay-substitution are given. Refer figure 9 in paper for B-I, B-II and B-III.
In left panels black circles show lowest singlet states and blue stars show lowest triplet states corresponding to substitution strength $\epsilon = 1, 2, 3$. In right panels red squares and green upper traingles represent first and second excited singlet states whereas magenta lower traingles and violet diamonds symbolize the first and second excited triplet states.}
\label{fig: sub-bay}
\end{figure}
\begin{table}[]
\caption{Energy gaps and corresponding transition dipole moments are given for $\epsilon=1$ in singlet and triplet states. In this table $S_n$ represents singlet states and $T_n$ represents various triplet states. Energy gaps are calculated against ground state of the system whereas transition dipole moments are with respect to the corresponding lowest state.}
\begin{tabular}{|c|c|c|c|c|c|c|c|c|}
\hline
\multicolumn{9}{|c|}{{\ul \textbf{Energies and dipole moments for $\epsilon =1$ }}} \\ \hline                                                                                                                                                                        
System     & State     & Energy     & $\mu_x$   &  $\mu_y$      & State    & Energy    &  $\mu_x$     &  $\mu_y$ \\ \hline \hline
\multirow{4}{*}{P-I}       & $S_1$        & 0.00     & 0.17     & 0.00     & $T_1$       & 1.53    & 0.79    & 0.00\\ \cline{2-9} 
                           & $S_2$        & 1.56     & 0.00     & 0.00     & $T_2$       & 2.94    & 0.00     & 0.00\\ \cline{2-9} 
                           & $S_3$        & 2.95     & 0.00     & 0.00     & $T_3$       & 2.96    & 0.00     & 0.00   \\ \cline{2-9} 
                           & $S_4$        & 2.98     & 0.00     & 0.00     & $T_4$       & 3.36    & 0.00     & 0.17  \\ \hline
\multirow{4}{*}{P-II}      & $S_1$        & 0.00     & 0.00     & 0.00     & $T_1$       & 1.56    & 0.00     & 0.00   \\ \cline{2-9} 
                           & $S_2$        & 1.59     & 0.00     & 0.00     & $T_2$       & 2.96    & 0.60    & 0.17  \\
\hline
\end{tabular}
\label{fig: trdip-e1}
\end{table}
\begin{table}[]
\caption{Bond orders in substituted perylenes for $\epsilon=1$. Bonds are given in first coulumn and according to figure 1.}
\begin{tabular}{|l|c|c|c|c|c|c|c|c|c|c|c|c|}
\hline
\multicolumn{13}{|c|}{{\ul \textbf{Bond Orders in lowest singlet state for $\epsilon = 1$}}}                                                                                                                                                                                                                                                     \\ \hline
Bonds/System & \multicolumn{2}{c|}{\textit{P-I}}    & \multicolumn{2}{c|}{\textit{P-II}}  & \multicolumn{2}{c|}{\textit{P-III}}                            & \multicolumn{2}{c|}{\textit{B-I}}   & \multicolumn{2}{c|}{\textit{B-II}}            & \multicolumn{2}{c|}{\textit{B-III}}                            \\ \hline
   & \multicolumn{1}{l|}{\textbf{Ms=0}} & \multicolumn{1}{l|}{\textbf{Ms=1}} & \multicolumn{1}{l|}{\textbf{Ms=0}} & \multicolumn{1}{l|}{\textbf{Ms=1}} & \multicolumn{1}{l|}{\textbf{Ms=0}} & \multicolumn{1}{l|}{\textbf{Ms=1}} & \multicolumn{1}{l|}{\textbf{Ms=0}} & \multicolumn{1}{l|}{\textbf{Ms=1}} & \multicolumn{1}{l|}{\textbf{Ms=0}} & \multicolumn{1}{l|}{\textbf{Ms=1}} & \multicolumn{1}{l|}{\textbf{Ms=0}} & \multicolumn{1}{l|}{\textbf{Ms=1}} \\ \hline \hline
1-2          & 0.69                      & 0.59                      & 0.69                      & 0.59                      & 0.67                      & 0.61                      & 0.70                      & 0.60                      & 0.70                      & 0.60                      & 0.67                      & 0.62                      \\ \hline
2-3          & 0.52                      & 0.52                      & 0.51                      & 0.53                      & 0.54                      & 0.56                      & 0.52                      & 0.53                      & 0.52                      & 0.53                      & 0.53                      & 0.55                      \\
\hline
\end{tabular}
\label{fig: BO-e1}
\end{table}
\begin{table}[]
\caption{Charge densities in substituted perylenes for $\epsilon=1$.}
\begin{tabular}{|l|c|c|c|c|c|c|c|c|c|c|c|c|}
\hline
\multicolumn{13}{|c|}{{\ul \textbf{Charge Densities in lowest singlet state for $\epsilon=1$}}}                                                                                                                                                                                                                                                                                                                                                                                     \\ \hline
\textit{Bonds/System} & \multicolumn{2}{c|}{\textit{P-I}}                                       & \multicolumn{2}{c|}{\textit{P-II}}                                      & \multicolumn{2}{c|}{\textit{P-III}}                                     & \multicolumn{2}{c|}{\textit{B-I}}                                       & \multicolumn{2}{c|}{\textit{B-II}}                                      & \multicolumn{2}{c|}{\textit{B-III}}                                     \\ \hline
                      & \multicolumn{1}{l|}{\textbf{Ms=0}} & \multicolumn{1}{l|}{\textbf{Ms=1}} & \multicolumn{1}{l|}{\textbf{Ms=0}} & \multicolumn{1}{l|}{\textbf{Ms=1}} & \multicolumn{1}{l|}{\textbf{Ms=0}} & \multicolumn{1}{l|}{\textbf{Ms=1}} & \multicolumn{1}{l|}{\textbf{Ms=0}} & \multicolumn{1}{l|}{\textbf{Ms=1}} & \multicolumn{1}{l|}{\textbf{Ms=0}} & \multicolumn{1}{l|}{\textbf{Ms=1}} & \multicolumn{1}{l|}{\textbf{Ms=0}} & \multicolumn{1}{l|}{\textbf{Ms=1}} \\ \hline \hline
1                     & 0.06                               & 0.05                               & 0.06                               & 0.05                               & 0.07                               & 0.06                               & 0.04                               & 0.06                               & 0.05                               & 0.06                               & 0.06                               & 0.05                               \\
\hline
\end{tabular}
\label{fig: chrden-e1}
\end{table}
\begin{table}[]
\caption{Spin densities in substituted perylenes for $\epsilon=1, 3$.}
\begin{tabular}{|l|c|c|c|c|c|c|c|c|c|c|c|c|}
\hline
\multicolumn{13}{|c|}{{\ul \textbf{Spin Densities in lowest triplet state for $\epsilon = 1, 3$}}}                                                                                                                                                                                                                                                                                                                                                                            \\ \hline
\textit{Bonds/System} & \multicolumn{2}{c|}{\textit{P-I}}                    & \multicolumn{2}{c|}{\textit{P-II}}                                        & \multicolumn{2}{c|}{\textit{P-III}}                                       & \multicolumn{2}{c|}{\textit{B-I}}                                         & \multicolumn{2}{c|}{\textit{B-II}}                                        & \multicolumn{2}{c|}{\textit{B-III}}                                       \\ \hline
                      & \textbf{$\epsilon=1$} & \multicolumn{1}{l|}{\textbf{$\epsilon=3$}} & \multicolumn{1}{l|}{\textbf{$\epsilon=1$}} & \multicolumn{1}{l|}{\textbf{$\epsilon=3$}} & \multicolumn{1}{l|}{\textbf{$\epsilon=1$}} & \multicolumn{1}{l|}{\textbf{$\epsilon=3$}} & \multicolumn{1}{l|}{\textbf{$\epsilon=1$}} & \multicolumn{1}{l|}{\textbf{$\epsilon=3$}} & \multicolumn{1}{l|}{\textbf{$\epsilon=1$}} & \multicolumn{1}{l|}{\textbf{$\epsilon=3$}} & \multicolumn{1}{l|}{\textbf{$\epsilon=1$}} & \multicolumn{1}{l|}{\textbf{$\epsilon=3$}} \\ \hline
1                     & -0.02          & -0.01                               & -0.02                               & -0.01                               & -0.018                              & -0.02                              & -0.02                               & -0.00                               & -0.02                               & 0.01                                & -0.02                            & -0.002                              \\
\hline
\end{tabular}
\label{fig: spnden-e13}
\end{table}
\begin{table}[]
\caption{Energy gaps and corresponding transition dipole moments are given for $\epsilon=3$ in singlet and triplet states. In this table $S_n$ represents singlet states and $T_n$ represents various triplet states. Energy gaps are calculated against ground state of the system whereas transition dipole moments are with respect to the corresponding lowest state.}
\begin{tabular}{|c|c|c|c|c|c|c|c|c|}
\hline
\multicolumn{9}{|c|}{{\ul \textbf{Energies and dipole moments for $\epsilon =3$ }}} \\ \hline                                                                                                                                                                        
System     & State     & Energy     & $\mu_x$   &  $\mu_y$      & State    & Energy    &  $\mu_x$     &  $\mu_y$ \\ \hline \hline
\multirow{4}{*}{P-I}       & $S_1$        & 0.00     & 0.46      & 0.00     & $T_1$       & 1.39    & 2.89    & 0.00\\ \cline{2-9} 
                           & $S_2$        & 1.41     & 0.00      & 0.00     & $T_2$       & 2.83    & 0.00     & 0.00\\ \cline{2-9} 
                           & $S_3$        & 2.36     & 0.00      & 4.52     & $T_3$       & 3.04    & 0.00     & 0.00   \\ \cline{2-9} 
                           & $S_4$        & 2.84     & 0.00      & 0.00     & $T_4$       & 3.20    & 0.00     & 0.73  \\ \hline
\multirow{4}{*}{P-II}      & $S_1$        & 0.00     & 0.00      & 0.00     & $T_1$       & 1.64    & 0.00     & 0.00   \\ \cline{2-9} 
                           & $S_2$        & 1.66     & 0.00      & 0.00     & $T_2$       & 2.98    & 0.39    & 1.32  \\
\hline
\end{tabular}
\label{fig: trdip-e3}
\end{table}
\begin{table}[]
\caption{Bond orders in substituted perylenes for $\epsilon=3$. Bonds are given in first coulumn and according to figure 1.}
\begin{tabular}{|l|c|c|c|c|c|c|c|c|c|c|c|c|}
\hline
\multicolumn{13}{|c|}{{\ul \textbf{Bond Orders in lowest singlet state for $\epsilon = 3$}}}                                                                                                                                                                                                                                                                                                                                                                                         \\ \hline
\textit{Bonds/System} & \multicolumn{2}{c|}{\textit{P-I}}                                       & \multicolumn{2}{c|}{\textit{P-II}}                                      & \multicolumn{2}{c|}{\textit{P-III}}                                     & \multicolumn{2}{c|}{\textit{B-I}}                                       & \multicolumn{2}{c|}{\textit{B-II}}                                      & \multicolumn{2}{c|}{\textit{B-III}}                                     \\ \hline
                      & \multicolumn{1}{l|}{\textbf{Ms=0}} & \multicolumn{1}{l|}{\textbf{Ms=1}} & \multicolumn{1}{l|}{\textbf{Ms=0}} & \multicolumn{1}{l|}{\textbf{Ms=1}} & \multicolumn{1}{l|}{\textbf{Ms=0}} & \multicolumn{1}{l|}{\textbf{Ms=1}} & \multicolumn{1}{l|}{\textbf{Ms=0}} & \multicolumn{1}{l|}{\textbf{Ms=1}} & \multicolumn{1}{l|}{\textbf{Ms=0}} & \multicolumn{1}{l|}{\textbf{Ms=1}} & \multicolumn{1}{l|}{\textbf{Ms=0}} & \multicolumn{1}{l|}{\textbf{Ms=1}} \\ \hline
1-2                   & 0.66                               & 0.57                               & 0.66                               & 0.57                               & 0.68                               & 0.60                               & 0.69                               & 0.60                               & 0.71                               & 0.61                               & 0.71                               & 0.59                               \\
\hline
\end{tabular}
\label{fig: BO-e3}
\end{table}
\begin{table}[]
\caption{Charge densities in substituted perylenes for $\epsilon=3$.}
\begin{tabular}{|l|c|c|c|c|c|c|c|c|c|c|c|c|}
\hline
\multicolumn{13}{|c|}{{\ul \textbf{Charge Densities in lowest singlet state for epsilon =3}}}                                                                                                                                                                                                                                                                                                                                                                                     \\ \hline
\textit{Bonds/System} & \multicolumn{2}{c|}{\textit{P-I}}                                       & \multicolumn{2}{c|}{\textit{P-II}}                                      & \multicolumn{2}{c|}{\textit{P-III}}                                     & \multicolumn{2}{c|}{\textit{B-I}}                                       & \multicolumn{2}{c|}{\textit{B-II}}                                      & \multicolumn{2}{c|}{\textit{B-III}}                                     \\ \hline
                      & \multicolumn{1}{l|}{\textbf{Ms=0}} & \multicolumn{1}{l|}{\textbf{Ms=1}} & \multicolumn{1}{l|}{\textbf{Ms=0}} & \multicolumn{1}{l|}{\textbf{Ms=1}} & \multicolumn{1}{l|}{\textbf{Ms=0}} & \multicolumn{1}{l|}{\textbf{Ms=1}} & \multicolumn{1}{l|}{\textbf{Ms=0}} & \multicolumn{1}{l|}{\textbf{Ms=1}} & \multicolumn{1}{l|}{\textbf{Ms=0}} & \multicolumn{1}{l|}{\textbf{Ms=1}} & \multicolumn{1}{l|}{\textbf{Ms=0}} & \multicolumn{1}{l|}{\textbf{Ms=1}} \\ \hline
1                     & 0.18                               & 0.15                               & 0.19                               & 0.15                               & 0.08                               & 0.07                               & 0.14                               & 0.19                               & 0.14                               & 0.20                               & 0.19                               & 0.17                               \\
\hline
\end{tabular}
\label{fig: chrden-e3}
\end{table}